%% file: main.tex
\documentclass[%
 reprint,
 amsmath,amssymb,
 aps,
 prx,
]{revtex4-2}

\usepackage{silence}
\usepackage{graphicx}
\usepackage{dcolumn}
\usepackage{bm}

\usepackage{amsmath}
\usepackage{amssymb}
\usepackage{mathtools}
\usepackage{amsthm}
\usepackage{newunicodechar}
\usepackage{xcolor}
\usepackage{tcolorbox}
\usepackage{needspace}

\usepackage{quantikz}
\usepackage{adjustbox}
\usepackage{hyperref}

\input{code_style}

\makeatletter
\providecommand{\fnum@lstlisting}{}
\renewcommand{\fnum@lstlisting}{\lstlistingname~\thelstlisting}
\newcommand{\listingcaption}[1]{%
  \refstepcounter{lstlisting}%
  \begingroup
    \@makecaption{\fnum@lstlisting}{#1}\par
  \endgroup
}
\makeatother

\begin{document}

\preprint{APS/123-QED}

\title{Automating quantum feature map design via large language models} 

\author{Kenya Sakka\textsuperscript{1}}
\author{Kosuke Mitarai\textsuperscript{1, 2}}
\author{Keisuke Fujii\textsuperscript{1, 2, 3}}

\affiliation{\textsuperscript{1}Center for Quantum Information and Quantum Biology,
  The University of Osaka, 1-2 Machikaneyama, Toyonaka 560-0043, Japan}
\affiliation{\textsuperscript{2}Graduate School of Engineering Science, The University of Osaka
1-3 Machikaneyama, Toyonaka, Osaka 560-8531, Japan}
\affiliation{\textsuperscript{3}RIKEN Center for Quantum Computing, Wako Saitama 351-0198, Japan}

\vspace{5pt}
\date{\today}

\begin{abstract}
Quantum feature maps are a key component of quantum machine learning, encoding classical data into quantum states to exploit the expressive power of high-dimensional Hilbert spaces.
Despite their theoretical promise, designing quantum feature maps that offer practical advantages over classical methods remains an open challenge.
In this work, we propose an agentic system that autonomously generates, evaluates, and refines quantum feature maps using large language models.
The system consists of five components: Generation, Storage, Validation, Evaluation, and Review.
Using these components, it iteratively improves quantum feature maps. 
Through numerical evaluations on widely used benchmark datasets, the system discovers and improves quantum feature maps without human intervention.
On MNIST, the best generated feature map achieves 97.3\% classification accuracy, outperforming existing quantum feature maps and achieving competitive performance with classical kernels, remaining within 0.3 percentage points of the radial basis function kernel.
Similar improvements are observed on Fashion-MNIST and CIFAR-10.
These results demonstrate that LLM-driven closed-loop discovery can autonomously explore dataset-adaptive quantum features.
More broadly, our approach provides a practical methodology for automated discovery in quantum circuit design, helping bridge the gap between theoretical QML models and their empirical performance on real-world machine learning tasks.
\end{abstract}

\maketitle

\section{Introduction}
\label{introduction}

\begin{figure*}[t]
    \includegraphics[width=1.0\linewidth]{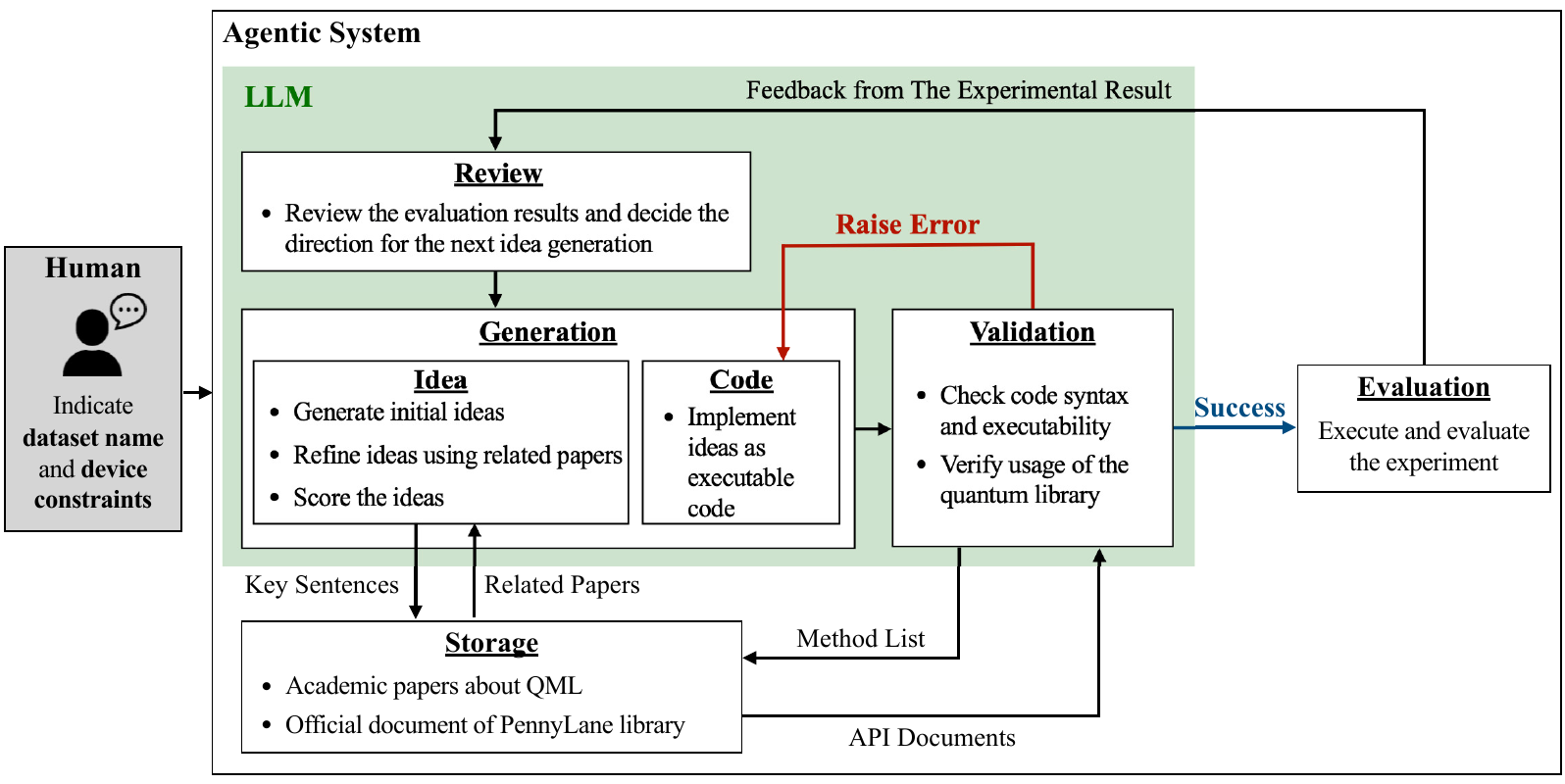}
    \caption{Overview of the agentic system for automatic generation of quantum feature maps. When a user provides task instructions to the system, five internal components work collaboratively to autonomously conduct numerical evaluations and improvements. As a result, the system generates an executable program that implements a quantum feature map capable of performing the task with high accuracy.}
    \label{tab:architecture}
\end{figure*}

Quantum machine learning (QML) has gained attention in recent years due to its potential advantages over classical machine learning~\cite{biamonte2017quantum, cerezo2022challenges}.
By leveraging the principles of quantum mechanics, QML aims to enhance computational efficiency and improve learning performance across various tasks.

A central concept in QML is that of quantum features~\cite{havlivcek2019supervised, schuld2019quantum, mitarai2018quantum, cerezo2022challenges}. 
This notion can be seen as a quantum counterpart of classical features in machine learning, where quantum states—represented as density operators on Hilbert spaces whose dimension grows exponentially with the number of qubits—are used as feature representations. 
By encoding classical data into such quantum states and leveraging the structure of these exponentially large Hilbert spaces, quantum features aim to provide enhanced expressive power for learning algorithms.
Liu \textit{et al.} have shown that, in a certain artificial task, quantum features can indeed offer rigorous advantages over classical counterparts~\cite{liu2021rigorous}.

Despite the theoretical promise of quantum features, designing quantum feature maps that provide practical advantages in real-world machine learning tasks remains an open challenge.
In particular, for widely used datasets like MNIST and other benchmark tasks in classical machine learning, no quantum feature map has yet been found that consistently outperforms classical approaches or demonstrates a clear quantum advantage.
This limitation has been highlighted in a recent study by Huang \textit{et al.}~\cite{huang2021power}, where various empirical quantum feature maps have been tested across standard benchmarks.
The results so far suggest that quantum advantages are hard to observe under realistic conditions, pointing to a gap between theoretical potential and practical applicability.

One fundamental reason for this difficulty is that, in both classical and quantum settings, effective feature maps are often highly dependent on the structure of the dataset and the nature of the task.
For instance, features suitable for image data may differ substantially from those effective for time-series or tabular data.
Therefore, the practical realization of useful quantum features must consider dataset-specific design, ideally allowing for automated adaptation to the data at hand.
Developing methods to generate such dataset-adaptive quantum features in a principled and scalable manner is an important and urgent challenge.
Addressing this challenge is critical for advancing QML from a theoretical concept to a practical tool in modern machine learning.

Research on AI-assisted quantum circuit design has developed in several directions, including quantum architecture search (QAS)~\cite{zhang2022differentiable, martyniuk2024quantum}, which aims to automatically discover task-specific circuit structures by exploring a predefined architecture space.
The use of Transformer~\cite{vaswani2017attention} based generative models, represented by large language models (LLMs), in quantum circuit design is still limited, but recent work has begun to reveal the potential of automation in quantum algorithm development.
For example, Nakaji~\cite{nakaji2024gqe} demonstrated an approach tailored to ground-state search problems, highlighting the feasibility of circuit generation by training task-specific models and validating the results through numerical evaluations.
In contrast, Ueda~\cite{Ueda2024Optimizing} proposed a framework that leverages LLMs to select appropriate ansatz and incorporate numerical evaluation results as feedback for the LLMs. 
However, this work focuses primarily on conceptual design and does not include implementation results or quantitative evaluation.
Nevertheless, these methods rely on fixed circuit templates, task-specific formulations that require additional model training or fine-tuning, and static internal knowledge, which can limit their flexibility and scalability across diverse tasks and rapidly evolving quantum software environments~\cite{basit2025pennylang}.
While such constraints stabilize exploration, they also restrict the accessible hypothesis space and embed architectural priors that may not generalize across datasets. 
In particular, realizing automated scientific discovery in quantum circuit design requires more than simple circuit generation.
Removing predefined templates dramatically enlarges the admissible hypothesis space, potentially destabilizing exploration.
At the same time, evaluating quantum feature maps involves substantially more than circuit construction, including device configuration and machine learning pipeline implementation, resulting in complex generated code.
Moreover, the performance of quantum feature maps is difficult to predict analytically, making empirical, trial-and-error refinement unavoidable.
As a consequence, coherently integrating code generation, executable validation, and iterative performance-based improvement remains a nontrivial open challenge.

In this work, we address these challenges by proposing a prompt-based system for autonomous improvement of quantum feature map design.
As a prompt-based system, it does not require any additional model training or fine-tuning, making it easily adaptable to various tasks.
Our system consists of five components—Generation, Storage, Validation, Evaluation, and Review—that together implement a structured closed-loop architecture integrating template-free circuit generation with executable validation and performance-based refinement.
An overview of the system developed in this work is shown in Fig. ~\ref{tab:architecture}.
To stabilize exploration in the absence of predefined templates, the Storage component incorporates a curated database of up-to-date quantum machine learning academic papers, providing structured scientific guidance beyond the LLM’s internal knowledge.
Implementation reliability is ensured through dedicated validation procedures prior to empirical evaluation, and candidate feature maps are iteratively refined based on dataset-level performance rather than one-shot generation.
To the best of our knowledge, this is the first work in the quantum domain to automate the entire research workflow, ranging from idea generation to implementation, evaluation, and iterative refinement, using LLMs.

Through numerical evaluations on widely used benchmark datasets, we demonstrate the effectiveness of the proposed system. 
On MNIST, the automatically generated feature maps achieve a test accuracy of 97.3\%, outperforming classical linear and polynomial kernels while remaining within 0.3 percentage points of the widely used radial basis function (RBF) kernel.
Similar improvements are observed on Fashion-MNIST and CIFAR-10, indicating that the discovered feature maps generalize across different datasets.
In QML, one of the central goals is to surpass the performance of classical machine learning models. 
In line with this objective, our system not only improves upon commonly used quantum feature maps but also competes with strong classical kernels.
Additional comparisons show that the generated feature maps improve accuracy by 2–5 percentage points over those discovered by existing QAS methods.
Ablation studies further validate the effectiveness of the proposed system architecture. 
By integrating LLM-driven hypothesis generation with executable quantum program validation and performance-based refinement, our work establishes a practical methodology for autonomous exploration of quantum feature representations, helping bridge the gap between theoretical QML models and their practical deployment.

\section{Preliminary}
This section outlines the fundamental concepts necessary for the subsequent discussion. We begin by introducing the concept of quantum feature maps, which encode classical data into quantum states. We then describe the quantum kernel method, where similarities between quantum states are used for learning tasks. Following this, we provide a brief overview of LLMs, focusing on their generative capabilities and broad applicability. Finally, we discuss recent developments in automated science, highlighting the role of LLMs in automating various stages of the scientific discovery process, including applications in quantum computing.

\subsection{Quantum Feature Map}
Feature mapping is a transformation $\phi: \mathcal{X} \to \mathcal{F}$ that maps data from the original input space $\mathcal{X}$ to a higher-dimensional feature space $\mathcal{F}$, facilitating the identification of nonlinear patterns. 
For instance, given an input $d$-dimensional vector $ \mathbf{x} \in \mathbb{R}^d $, an appropriately designed nonlinear transformation $\phi(\mathbf{x})$ can allow linear models, such as support vector machines (SVMs), to effectively handle complex data distributions.
However, increasing the complexity of the nonlinear transformation generally requires a higher-dimensional feature space, which leads to the curse of dimensionality. Kernel methods address this challenge by enabling computations in the high-dimensional space without explicitly constructing it, using a kernel function that computes inner products between mapped data points. This concept forms the basis of kernel-based learning algorithms and naturally motivates their extension to quantum settings.

Quantum feature mapping extends this concept to quantum computing by employing a quantum circuit to encode classical data into quantum states~\cite{havlivcek2019supervised, schuld2019quantum, mitarai2018quantum, cerezo2022challenges}. Given an input $ \mathbf{x} $, the quantum feature map prepares the corresponding quantum state, or quantum feature, as
\begin{equation}
\rho(\mathbf{x}) = U(\mathbf{x})|0\rangle^{\otimes n}\langle0|^{\otimes n} U(\mathbf{x})^\dagger,
\end{equation}
where $ |0\rangle^{\otimes n} $ is the initial state of $ n $ qubits, and $ U(\mathbf{x}) $ is a unitary transformation that depends on $ \mathbf{x} $.

\subsection{Quantum Kernel Method}
The quantum kernel method applies classical kernel techniques to data embedded in a quantum Hilbert space via quantum feature maps.
The quantum kernel function is defined as the Hilbert–Schmidt inner product between density operators:
\begin{equation}
k(\mathbf{x}, \mathbf{x}') = \operatorname{Tr}[\rho(\mathbf{x}) \rho(\mathbf{x}')] \label{eq:kernel}
\end{equation}
This kernel function quantifies the similarity between data points in the quantum feature space and can be used with standard kernel-based learning algorithms such as support vector machines or kernel ridge regression~\cite{havlivcek2019supervised, schuld2019quantum, cerezo2022challenges}.

An attractive property of the quantum kernel method is that it inherits the theoretical foundation of classical kernel methods, enabling one to find optimal solutions in the quantum feature space according to a well-defined loss function. In particular, learning remains convex in the feature space, allowing for efficient optimization and generalization analysis~\cite{havlivcek2019supervised, schuld2019quantum, cerezo2022challenges}. 
This is in contrast to other approaches that employ the quantum feature to construct models, such as quantum circuit learning framework~\cite{mitarai2018quantum} where we apply a trainable quantum circuit to $\rho(\mathbf{x})$ to extract relevant information from the feature.
These approaches gives lower prediction cost~\cite{nakayama2024explicit} and might be able to generalize more due to the restricted search space~\cite{mitarai2018quantum}, the complex loss landscape challenges us to train the models~\cite{mcclean2018barren}.
As such, researchers have often employed the quantum kernel methods to benchmark the quantum feature maps on real datasets~\cite{huang2021power, haug2021quantum}.

\subsection{Large Language Models}
Large Language Models (LLMs) are a type of generative model and large-scale machine learning system trained on vast amounts of text data from the internet and other sources~\cite{anthropic2024claude3, google2023gemini, llama3herd2024, openai2023gpt4}. Given an input sequence of tokens $ x_{1:t} = (x_1, x_2, \dots, x_t) $, LLMs generate text by modeling the conditional probability distribution:
\begin{equation}
P(x_{t+1} \mid x_{1:t}) = \frac{\exp(s(x_{1:t}, x_{t+1}))}{\sum_{x' \in V} \exp(s(x_{1:t}, x'))}
\end{equation}
where $ s(x_{1:t}, x') $ represents the unnormalized logit score assigned to token $ x' $, and $ V $ denotes the vocabulary. By sampling from this distribution, LLMs iteratively produce coherent and contextually relevant text.

Leveraging extensive training datasets and large-scale neural networks, such as the Transformer architecture~\cite{vaswani2017attention}, LLMs can perform not only text generation but also a wide range of tasks, including code generation~\cite{li2022alphacode} and multi-modal processing involving images, videos, and audio~\cite{yin2024survey}. Furthermore, in recent years, models referred to as Reasoning Models~\cite{huang2023towards, deepseek2025deepseekr1}, which are designed for deep and structured thinking, have been increasingly utilized in tasks requiring complex task.

\subsection{Automated Science}
The use of AI for scientific discovery has primarily been explored in domains that can be simulated, such as machine learning. In particular, the recent significant advancements in LLMs have greatly expanded their applicability, as computers can now understand instructions provided in natural language. For example, there have been efforts to use LLMs for generating objective functions in machine learning models~\cite{lu2024discopop}. In that research, LLMs were employed to automate the generation of objective functions and their subsequent refinement based on evaluation results. Additionally, research has focused on model merging algorithms, proposing an iterative improvement process driven by LLMs to develop algorithms autonomously, without relying on human expertise or predefined ideas~\cite{ishibashi2024selfdeveloping}. In addition to work on automating specific algorithm development, there has also been research on the automation of scientific discovery in collaboration with humans~\cite{qian2024towards}, aiming to support and enhance human-led research processes. Beyond this, other studies have explored the complete automation of the scientific research process, encompassing idea generation, execution, and academic paper writing~\cite{lu2024aiscientist, yamada2025aiscientistv2}.

Regarding the use of AI in the field of quantum computing, generative models based on Transformer architectures have been developed to generate quantum circuits~\cite{nakaji2024gqe}. 
However, these models are primarily trained for the ground state search problem, and extending them to tasks in different domains requires the design of new objective functions and re-training, which can be both technically challenging and computationally expensive. 
Moreover, they cannot readily incorporate up-to-date external knowledge, and there is no established method for encoding classical data, limiting their applicability.
Also, Ueda and Matsuo have proposed to use LLMs for optimizing quantum circuits in quantum generative adversarial networks~\cite{Ueda2024Optimizing}. 
In their approach, the LLM selects from a set of ansatz candidates predefined by the user, rather than generating code directly.
The method depends on the model’s internal knowledge, and its practicality remains unclear, as no numerical evaluations or implementation results are provided—it is presented primarily as a conceptual idea.

The role of AI in scientific automation is expanding, evolving from a mere assistive tool to an entity that actively participates in the fundamental process of scientific discovery.
However, to truly realize this potential, new frameworks are needed to enable AI to engage efficiently in the pursuit of novel scientific insights.
This includes not only supporting individual tasks but also empowering AI to operate across the entire research cycle—generating idea, designing and executing experiments, analyzing results, and iteratively refining its own processes—to achieve autonomous scientific advancement.

\section{LLM-based system for automatic generation of quantum feature maps} 
Our agentic system for automatic generation of quantum feature maps consists of five components: ``Generation'', ``Storage'', ``Validation'', ``Evaluation'' and ``Review'' (Fig.~\ref{tab:architecture}). A single trial comprises processing by these five components, and by iteratively repeating this process while incorporating feedback from the results, the system improves the classification accuracy progressively.
In this section, we describe the architecture of our system in detail.

\subsection{Generation} \label{sec:methodsGeneration}

In the “Generation” component, the system uses LLMs to propose, score, refine, and implement candidate ideas for quantum feature maps.
This process begins with the generation of multiple candidate ideas, followed by a scoring phase to evaluate their potential. Based on the scores and other contextual signals, the system then performs a reflection step to refine ideas. 
These refined ideas are re-scored, and finally, the ideas are implemented as Python programs.
Each prompt includes information such as the dataset name, the kernel function, the machine learning model, the input data format, hardware constraints of the quantum device, and other directives. 

\subsubsection{Idea generation}

The first process of the ``Generation'' component is generation of candidate ideas.
Here, the system prompts an LLM to generate ideas for quantum feature maps.
For the first trial, we designed a prompt that encourages the generation of broadly applicable, general ideas to create a foundation for future performance improvements.
In the subsequent trials, the system prompt LLM to refine the design of the ansatz based on the review provided from the previous Review step.
Throughout the trials, we prompt LLM to generate multiple ideas simultaneously to encourage diversity.
Notably, we prohibit the use of nonlinear transformation or trainable parameters within the quantum feature maps, since otherwise the system might output feature maps that depend on classical models, such as neural networks, to achieve high accuracy.
The LLM structures each output into four components: an overview, a detailed explanation, corresponding mathematical expressions in TeX format, and a set of key sentences summarizing the idea.
As examples, we provide the prompts used for this idea generation process in this work (see Sec.~\ref{sec:evaluationSetup}) in Listing~\ref{lst:IdeaGenerationDeveloper}, \ref{lst:IdeaGenerationUserFirst} and \ref{lst:IdeaGenerationUserSubsequent} in Appendix~\ref{sec:prompts}.

\subsubsection{Scoring}\label{sec:scoring}

The second process of the ``Generation'' is scoring of the generated ideas.
The system prompts an LLM to score the overall direction of each idea based on three criteria: \textit{Originality}, which assesses the novelty of the idea; \textit{Feasibility}, which examines whether the idea can be implemented as a program; and \textit{Versatility}, which evaluates broad applicability of the idea rather than being overly specialized for a particular task or dataset.
For this step, we integrate a vector database storing relevant information within the system, which is described in detail in Sec.~\ref{sec:methodsStorage}.
It is notable that we assign the scores to the ideas solely to guide the LLM's reasoning and do not explicitly use them in later stages to, e.g., select which direction to persue. 
We eventually pass these scores to the user prompt of reflection for further refinement (Listing~\ref{lst:IdeaReflectionUser}).

The detailed flow of the scoring process is as follows.
The system first uses the key sentences generated during the idea generation phase as search queries to retrieve relevant academic papers from the database.
The LLM generate additional search queries up to a fixed number of attempts if it judges the initial information to be insufficient.
Note that naively inputting academic papers into the LLM can easily exceed its context window.
To avoid this, we use a lightweight model to generate concise summaries that fit within a predefined word limit, focusing on key elements such as methodology, results, and areas for improvement.
We provide the prompt used in the summarization process of this work in Listing~\ref{lst:SummarizationDeveloper} and \ref{lst:SummarizationUser}, in Appendix~\ref{sec:prompts}. 
The technique of retrieving necessary external information and incorporating it into the prompt of a generative model is known as Retrieval-Augmented Generation (RAG)~\cite{lewis2020retrieval}.
After the retrieval process is completed, the LLM scores the generated idea on a scale of 0 to 10 using the retrieved information.
To establish a scoring baseline, we also provide human-annotated scores of existing quantum feature maps to the LLM as few-shot examples~\cite{brown2020language}.
We use Listing~\ref{lst:IdeaScoringFewShots} in Appendix~\ref{sec:prompts} in our evaluations presented in \ref{sec:evaluationSetup}.
To reduce the variations of scores across trials, the scoring results are incorporated into the prompt for the next trial.
As examples, we provide the prompts used for this idea scoring process in this work (see Sec.~\ref{sec:evaluationSetup}) in Listing~\ref{lst:IdeaScoringDeveloper}, \ref{lst:IdeaScoringUserFirst}, \ref{lst:IdeaScoringUserIntermediate} and \ref{lst:IdeaScoringUserFinal}, in Appendix~\ref{sec:prompts}.

\subsubsection{Reflection}
The third process is to let an LLM to reflect on the generated ideas~\cite{madaan2023selfrefine}.
The system does so using external information, which are retrieved from the database using key sentences incorporated with the ideas as search queries. 
Concretely, the system first retrieve the summaries of the academic papers in the same manner during the scoring process, and then prompts an LLM to reflect on the ideas using them. 
This process of searching for relevant papers and reflecting on the idea for further improvements continues until the LLM judges the reflection to be complete or the predefined maximum number of iterations is reached.
As examples, we provide the prompts used for this reflection process in this work (see Sec.~\ref{sec:evaluationSetup}) in Listing~\ref{lst:IdeaReflectionDeveloper} and \ref{lst:IdeaReflectionUser}, in Appendix~\ref{sec:prompts}.

\subsubsection{Implementation}
Upon completion of idea reflection and scoring, the system finally uses an LLM to generate a Python program which implements quantum feature maps. 
In this process, the LLM is given the names of available quantum gates and library functions, as well as the implementation template presented in Listing~\ref{lst:BaseFeatureMapCode}. 
As examples, we provide the prompts used for this implementation process in this work (see Sec.~\ref{sec:evaluationSetup}) in Listing~\ref{lst:CodeGenerationDeveloper} and \ref{lst:CodeGenerationUser}, in Appendix~\ref{sec:prompts}.

\subsection{Storage} \label{sec:methodsStorage}
The ``Storage'' component of the system is a vector database consisting of relevant papers and documentations of the program libraries used by the LLMs.
The former is included to compensate for the knowledge cutoff of the LLMs, particularly in light of the rapid recent developments in quantum computing and quantum machine learning.
The latter addresses the need for up-to-date documentation, given the frequent updates in quantum computing software libraries.
This component stores these data as a vector database (VectorDB)~\cite{johnson2017billion, wang2021milvus}. A VectorDB accepts natural language sentences converted into vector representations as search queries, enabling context-aware retrieval that is more effective than traditional keyword-based search~\cite{karpukhin2020dense}.

\subsection{Validation} \label{sec:methodsValidation}

The ``Validation'' component of the system iteratively refine the code until it became executable by leveraging syntax analysis and external documentation.
It consists of three static validations and one dynamic validation.
As static validation, the system first checks whether the generated program can be compiled as a Python script using the \texttt{py\_compile} library.
Next, it uses the \texttt{ast} module to verify that the code complies with Python’s syntax rules.
If both checks pass, the system extracts PennyLane method names from the code and queries the documentation database using exact keyword matches.
The code and corresponding documentation are then provided as input to the LLM, which verifies the correctness of method usage, including function calls, argument names, and argument types.

After three static validations are passed, the system performs dynamic validation using dummy data to ensure that the code runs without errors.
As examples, we provide the prompts used for this validation process in this work (see Sec.~\ref{sec:evaluationSetup}) in Listing~\ref{lst:CodeValidationDeveloper} and \ref{lst:CodeValidationUser}, in Appendix~\ref{sec:prompts}.

If an error occurs at any stage of the validation process, the source code and error messages are fed back to the LLM, which attempts to correct and regenerate the source code. 
This validation process continues until all validations pass successfully or the maximum number of retries is exceeded. 
As examples, we provide the prompts used for this idea scoring process in this work (see Sec.~\ref{sec:evaluationSetup}) in Listing~\ref{lst:ErrorCorrectingDeveloper} and \ref{lst:ErrorCorrectingUser}, in Appendix~\ref{sec:prompts}.

\subsection{Evaluation} \label{sec:methodsEvaluation}
The ``Evaluation’’ component assesses the performance of the quantum machine learning model using validated quantum feature maps. It follows a structured procedure based on standard machine learning practices.

The system first splits the dataset into training, validation, and test subsets. 
It transforms the training samples using the generated quantum feature map and computes pairwise kernel values to construct a kernel matrix. Using this matrix, it trains a support vector machine (SVM). 
The trained SVM is then used to compute accuracy, precision, recall, and F-measure on the validation and test sets.
Note that the choice of the quantum-feature-based model is arbitrary; while we use an SVM in this work for demonstration purposes due to its ease of training once the kernel matrix is computed, other models such as quantum circuit learning~\cite{mitarai2018quantum} can also be employed.
After the evaluation, the system feeds back the results to ``Review'' component which guide idea generation in the next iteration.

\subsection{Review} \label{sec:methodsReview}
The ``Review'' component evaluates the quantum feature map ideas generated in the previous trial based on the feedback information provided by the ``Evaluation'' component.
This component uses an LLM to analyze the evaluation results of the most recent trial, listing multiple key factors that contributed to success and areas requiring improvement.
The prompt includes five key items: a text that guides the direction of the review, a textual description of the quantum feature map, its mathematical expression, the model’s training time, hardware information, and the performance metrics on the validation set, which includes an additional formatted string (e.g., idea\_1 $>$ idea\_2) to indicate the ranking of generated ideas based on accuracy.
The description and mathematical expression are the ones generated during the “Generation” step.
The results of this review are used as supplementary information for the ``Generation'' component in the next trial.

Four remarks are in order.
First, we include the training time in order to make the system to avoid quantum feature maps that are overly complex or rely on computationally expensive methods, such as amplitude encoding~\cite{schuld2015introduction}.
While such designs are not inherently undesirable from the perspective of achieving high accuracy, they can lead to prohibitively long simulation times in the numerical evaluations presented in Sec.~\ref{sec:evaluationSetup}.
For this reason, we encourage the system to account for computational cost.

Second, we include the hardware information in the prompt to prevent the system from deviating from the intended focus on quantum feature maps. 
In some cases, the generated ideas shifted toward broader challenges in quantum computing, such as noise reduction, while our main intension is to construct an effective quantum feature maps. 
For example, without this information, the ``Review’’ component occasionally produced suggestions to reduce the number of gates or to introduce redundant procedures for mitigating noise. 
While such ideas are relevant in general, they fall outside the scope of evaluating quantum feature maps. 
By explicitly specifying in the prompt that the evaluation is performed using a noiseless simulator, we reinforce that the evaluation should remain focused on the quantum feature map itself.

Third, we choose the text for directing the review with the following process.
The system first calculates the accuracy difference between the two most recent trials as $\mathrm{DiffMetric}$.
More specifically, $\mathrm{DiffMetric}$ at the $t$-th trial is calculated as $\mathrm{DiffMetric}(t) = A(t) - A(t-1)$, where $A(t)$ denotes the validation accuracy at the $t$-th trial.
Based on the $\mathrm{DiffMetric}$ value, it chooses five types of direction:
\begin{itemize}
\item $0.2 < \mathrm{DiffMetric} \leq 1.0$: The previous trial’s idea led to a significant improvement. The text tells the LLM to focus on identifying the successful aspects to ensure that the accuracy continues to improve.
\item $0.0 < \mathrm{DiffMetric} \leq 0.2$: The previous trial’s idea led to a moderate improvement. The text tells the LLM to identify the positive aspects while also exploring potential improvements for further accuracy enhancement.
\item $\mathrm{DiffMetric} = 0.0$: The previous trial’s idea resulted in no change. The text tells the LLM to identify bottlenecks or constraints in the idea and explore potential modifications.
\item $-0.2 \leq \mathrm{DiffMetric} < 0.0$: The previous trial’s idea led to a decline in accuracy. The text refers to all past trials and tells the LLM to determine the cause of the accuracy degradation and aim to restore the accuracy level.
\item $-1.0 \leq \mathrm{DiffMetric} < -0.2$: The previous trial’s idea led to a significant decline in accuracy. The text tells the LLM to conduct thorough analysis of all past trials, identify the root cause of accuracy deterioration, and consider major modifications to restore accuracy.
\end{itemize}

Finally, we direct the component to give a limited number of suggestions to the ``Generation'' module.
This design choice is based on the observation that incorporating many suggestions at simultaneously can lead to a lack of coherent direction in idea development, which may negatively affect accuracy.

We provide the prompts used for the Review component in this work (see Sec.~\ref{sec:evaluationSetup}) in Listing~\ref{lst:ReviewDeveloper} and \ref{lst:ReviewUser}, in Appendix~\ref{sec:prompts}.

\section{Numerical Evaluation Setup}\label{sec:evaluationSetup}

We numerically evaluate the effectiveness of our LLM-based agentic system for generating quantum feature maps on standard image classification datasets.
We first apply the system on the MNIST handwritten digits dataset~\cite{lecun1998mnist} to generate quantum feature maps.
Then, we compare the generated quantum feature maps with classical and quantum baselines on additional image datasets, specifically Fashion-MNIST~\cite{xiao2017fashion} and CIFAR-10~\cite{krizhevsky2009learning} to assess their generalizability.
In addition, we evaluate the diversity of the generated ideas, examine the scalability of the generated quantum feature map with respect to the number of qubits, and compare their accuracy with representative QAS approaches.
We also conduct an ablation study, in which we disable some of the components in the system to know their necessity.
The detailed evaluation protocol is described in the following sections.
Hyperparameters within the system, such as the number of ideas generated in each trial, can be found in Appendix~\ref{sec:hyperparameters}.

\subsection{Dataset and preprocessing}\label{sec:datasetPreprocessing}
The MNIST dataset~\cite{lecun1998mnist} consists of 70,000 handwritten digit images, each of which is a 28$\times$28 grayscale image labeled with one of the digits from 0 to 9. 
MNIST is officially divided into 60,000 training images and 10,000 test images. 
Our agentic system uses a sampled subset of the training data and does not access the official test set. 
When it finishes the iterative improvements of the feature maps, we evaluate the feature maps generated from the system on the full test dataset.
We also use the Fashion-MNIST~\cite{xiao2017fashion} and CIFAR-10~\cite{krizhevsky2009learning} datasets to evaluate the generalizability of the generated quantum feature map. 
Fashion-MNIST comprises 70,000 images related to fashion, each of which is a 28$\times$28 grayscale image categorized into one of ten classes, such as clothing and footwear. 
CIFAR-10 is a dataset consisting of 60,000 color images, where each image is a 32$\times$32 RGB image. 
The dataset includes ten classification labels, such as automobiles, birds, and other object categories. 

We apply principal component analysis (PCA) to reduce the dimensionality of the image data.
This is a standard approach when constructing quantum feature map for relatively high-dimensional inputs such as images~\cite{huang2021power}.
In this work, we use the top 80 principal components as input features and normalize the data to the range of 0.0 to 1.0.
This preprocessing is fixed, that is, the system never sees the original data at any point of the iterative process or allowed to modify this preprocessing.

To improve computational efficiency during the iterative quantum circuit generation process, we do not use the full official training set of 60,000 images within the MNIST dataset. 
Instead, we randomly sample 10,000 images while ensuring an equal distribution across all ten classes.
We then split this sampled dataset into 6,000 images for training, 2,000 for validation, and 2,000 for testing.

\subsection{Large language model}
We use OpenAI's LLMs for quantum feature map generation.
The system uses three models, ``o3-mini-2025-01-31'', ``gpt-4o-2024-11-20'', and ``gpt-4o-mini-2024-07-18'' depending on the specific tasks. 
Specifically, it uses the ``o3-mini-2025-01-31'' model, which excels in reasoning tasks, for review, idea-generation, idea-reflection, and code-generation. 
The \textit{reasoning\_effort} parameter, which controls the depth of reasoning, of this model is set to its highest level, ``high''. 
The parameter \textit{temperature}, which controls the randomness of output, is not supported in the "o3-mini-2025-01-31" model, and thus the output exhibits a fixed level of randomness.
The ``gpt-4o-2024-11-20'' model, a highly versatile general-purpose model, is utilized for scoring and validation tasks. The lightweight ``gpt-4o-mini-2024-07-18'' model is used for summarization tasks. 
For the GPT-4o series, the temperature parameter is set to 0.0, the lowest value indicating the least randomness.
We also perform evaluations using other LLMs, whose results are discussed in Appendix~\ref{sec:otherModels}.

When storing data in the VectorDB, we use the ``text-embedding-3-small'' model provided by OpenAI  as the embedding model. 
We segment texts into chunks of 1,024 tokens, and convert each chunk into a 1,536-dimensional vector representation before being stored in the database.

\subsection{External knowledge}
We use the arXiv API to retrieve academic papers in PDF format. We restrict the search criteria to the quant-ph category, which corresponds to quantum physics, and set the target period from January 1, 2020, to December 31, 2024. We specify “Quantum Machine Learning” as the search keyword, resulting in 998 retrieved papers.
To store them in the database, we extract text from the retrieved PDFs and segment it into chunks of 1,024 tokens.
We then convert these text segments into vector representations using an embedding model.

We use the source code of PennyLane version 0.39.0 as a reference for software documentation. 
Since PennyLane is open-source software, both its source code and documentation are publicly accessible.
However, the documentation is not always up-to-date and lacks sufficient detail for the purpose of this study.
To address this, we construct reference information for the LLM directly from the source code.
We begin by extracting all classes related to quantum gates, along with their class names and the corresponding docstrings that describe their functionality and usage.
We also add metadata that specifies how each class is invoked in a program.
Finally, we segment the docstring texts into chunks of 1,024 tokens and convert them into vector representations using an embedding model before storing them in the database.

\subsection{Support vector machine settings}
We assess the performance of the generated quantum feature maps using an SVM as the downstream model. 
For each feature map, we compute the kernel values using the Hilbert–Schmidt inner product defined in Eq.~\ref{eq:kernel}, and construct the corresponding kernel matrix. 
We then train an SVM classifier using \texttt{scikit-learn}~\cite{scikit-learn} based on this matrix. 
We choose the SVM for its efficiency and reproducibility in kernel-based learning tasks.

We fix the SVM hyperparameters as follows: we set the regularization parameter \texttt{C} to 1.0, and set the kernel coefficient \texttt{gamma} to \texttt{'scale'}, which automatically adjusts based on the variance of the input features. 
We perform all kernel evaluations using a noiseless simulator provided within PennyLane~\cite{Bergholm2018}. 
Throughout this study, the number of qubits is fixed to 10.

\subsection{Analysis beyond proof-of-concept}
In this subsection, we provide a more detailed analysis of the proposed system beyond a basic proof-of-concept on standard benchmarks. 
To characterize the properties of the automatically generated quantum feature maps, we consider four aspects: diversity of the generated feature maps, scalability with respect to the number of qubits, comparison with existing QAS approaches, and ablation studies of individual system components. 
The corresponding evaluation procedures are described below.

\subsubsection{Evaluating diversity of generated feature maps}
We quantify the diversity of the generated quantum feature maps by computing the Frobenius distance between their associated kernel matrices. 
We use a sampled subset of 6,000 MNIST training images, following the same data configuration adopted during feature map search by the agentic system.
For two ideas in the first trial, we compute the Frobenius distance between the kernel matrix of each of the 29 subsequently generated feature maps, 58 feature maps in total, and the kernel matrix associated with the corresponding initial idea.
We summarize these distances for each seed by reporting the mean, standard deviation, minimum, and maximum, which serve as indicators of the diversity of the agentic system's exploration range.
To assess structural dissimilarity between kernel matrices while controlling for their scale, we employ the normalized Frobenius distance. 
Given two kernel matrices $K_1$ and $K_2$, we first normalize each to unit Frobenius norm:
\begin{equation}
    \widetilde{K}_i = \frac{K_i}{\|K_i\|_F}, \quad \text{for } i = 1, 2,
\end{equation}
where $\|K\|_F$ denotes the Frobenius norm:
\begin{equation}
    \|K\|_F = \sqrt{ \sum_{i,j} K_{ij}^2 }.
\end{equation}
Then, the normalized Frobenius distance is defined as:
\begin{equation}
    d_F(\widetilde{K}_1, \widetilde{K}_2) = \left\| \widetilde{K}_1 - \widetilde{K}_2 \right\|_F.
\end{equation}
The value of $d_F$ lies in the interval $[0, 2]$, where $0$ indicates identical matrices, and values closer to $2$ reflect dissimilarity between kernel matrices.

\subsubsection{Scalability analysis}
We verify the scalability of the generated quantum feature map's performance.
Our agentic system explicitly specifies the target number of qubits in the LLM prompt.
As a result, some parts of the source code are implemented in a way that depends on the specified qubit count.
For evaluation, we manually introduce minimal adjustments to these qubit dependent parts while keeping the core idea unchanged, so that the circuits can run with different numbers of qubits. 
We also confirm that these adjustments do not alter the intended idea by verifying that the performance at ten qubits remains unchanged.
We then evaluate the modified circuits on the MNIST full test set for systems ranging from two to twelve qubits, in increments of two.

\subsubsection{Comparison with existing QAS approaches}
\label{sec:qasSettings}
We validate the effectiveness of our agentic system by comparing it with existing QAS approaches. 
We focus on two representative prior works by Lei \textit{et al.} \citep{lei2024neural} and Zhang \textit{et al.} \citep{zhang2021neural}, both of which target image classification.
Because the number of samples and other evaluation conditions in these studies differ from ours, we first align the conditions and then compare classification accuracy against the quantum feature map generated by our agentic system.

Lei's study~\cite{lei2024neural} uses a downsampled subset of MNIST consisting of digits 0 to 4 with 60 samples per class, for a total of 300 examples.
While details such as the random number seed and training test split strategy are not reported, we evaluate the performance of the generated feature map against theirs under two plausible settings: (i) using 300 samples for training and a separate disjoint set of 300 samples for testing, and (ii) splitting the same 300 samples into training and test sets with an 80 to 20 ratio.
For the prior study’s performance, we use the value reported in Appendix C.11 of Lei \citep{lei2024neural}, where the PCA reduced dimension of the input images matches that of our study. 

Zhang's study~\cite{zhang2021neural} focuses on Fashion-MNIST and considers a binary classification task between labels 0 (T shirt) and 3 (Dress), using 500 training and 500 test samples.
We replicate their evaluation protocol and assess our generated quantum feature map under the same conditions. 
For the prior study’s performance, we use the value reported in Fig.~7 of Ref.~\citep{zhang2021neural}.

\subsubsection{Ablation studies}
To assess the necessity and effectiveness of each component in the agentic system, we conduct ablation studies by systematically removing or modifying key components and evaluating their impact on system performance.
Each evaluation condition is tested over ten runs to account for the stochastic nature of the system, and the results are compared in terms of the mean and standard error of the best accuracy of test dataset achieved in each trial.

We first disable the ``Storage'' component.
Our system retrieves relevant academic papers stored in the VectorDB and provided them as contextual input to the LLM during the ``Scoring'' and ``Reflection'' components.
To evaluate the necessity of this process, we conducted a comparative analysis in which the LLM performed ``Scoring'' and ``Reflection'' solely based on its internal knowledge, without access to any external context.

Second, we modify the ``Review'' component.
For this ablation study, the modified components do not affect the execution or outcome of the first trial.
Accordingly, the system is initialized using the results and logs from Trial~1 of the default configuration, and the remaining trials are conducted from this shared initial state.
The ``Review'' component analyzes the performance metrics of each generated idea to enable iterative refinement of quantum feature maps.
As described in Section~\ref{sec:methodsReview}, the review process also incorporates instructions on the review direction into the prompt, based on the difference between the current trial accuracy and that of the immediately preceding trial.
To assess the necessity of this process, we conduct an ablation study under two conditions: one in which the direction determined from the $\mathrm{DiffMetric}$ is omitted from the prompt, and another in which the metric values from the current trial are directly passed to the ``Generation'' module without any review step.

\subsubsection{Effect of design parameter variations}
In addition to the ablation studies, we investigate the impact of key design parameters that control the behavior of the ``Review'' and ``Generation'' components.
In contrast to the preceding ablation studies, the following analyses do not remove system components but instead examine how varying these parameters influences overall performance.

We first vary the number of suggestions generated by the ``Review'' component.
As in the ablation study on the ``Review'' component, changing this parameter does not affect the execution or outcome of the first trial, and the system is therefore initialized using the results and logs from Trial~1 of the default configuration.
As described in ~\ref{sec:methodsReview}, the number of suggestions provided in a single review is intentionally limited.
In this analysis, we investigate the effect of the number of suggestions by varying it from one to three.
We additionally consider an ``auto'' setting, which corresponds to the prompt does not explicitly specify the number of suggestions.
In the default system configuration, the number of suggestions is set to three. 

Second, we vary the number of ideas generated in the ``Generation'' component.
Generating multiple ideas is expected to contribute not only to greater stability in accuracy but also to more diverse feedback for the ``Review'' component through performance evaluation.
In this analysis, we examine the impact of the number of generated ideas per trials on the overall system by varying it from one to three.
In the default system configuration, the number of ideas is set to two.

\section{Results}
\label{sec:results}
\input{tables/feature_map_performance} 

We first confirmed that our system can successfully generate executable quantum feature maps using an LLM. In all trials, the system generated Python code that passed the validation process without error, and we observed no execution failures after validation.

\subsection{Behavior of the improvements}

\begin{figure}[ht]
    \includegraphics[width=1.0\linewidth]{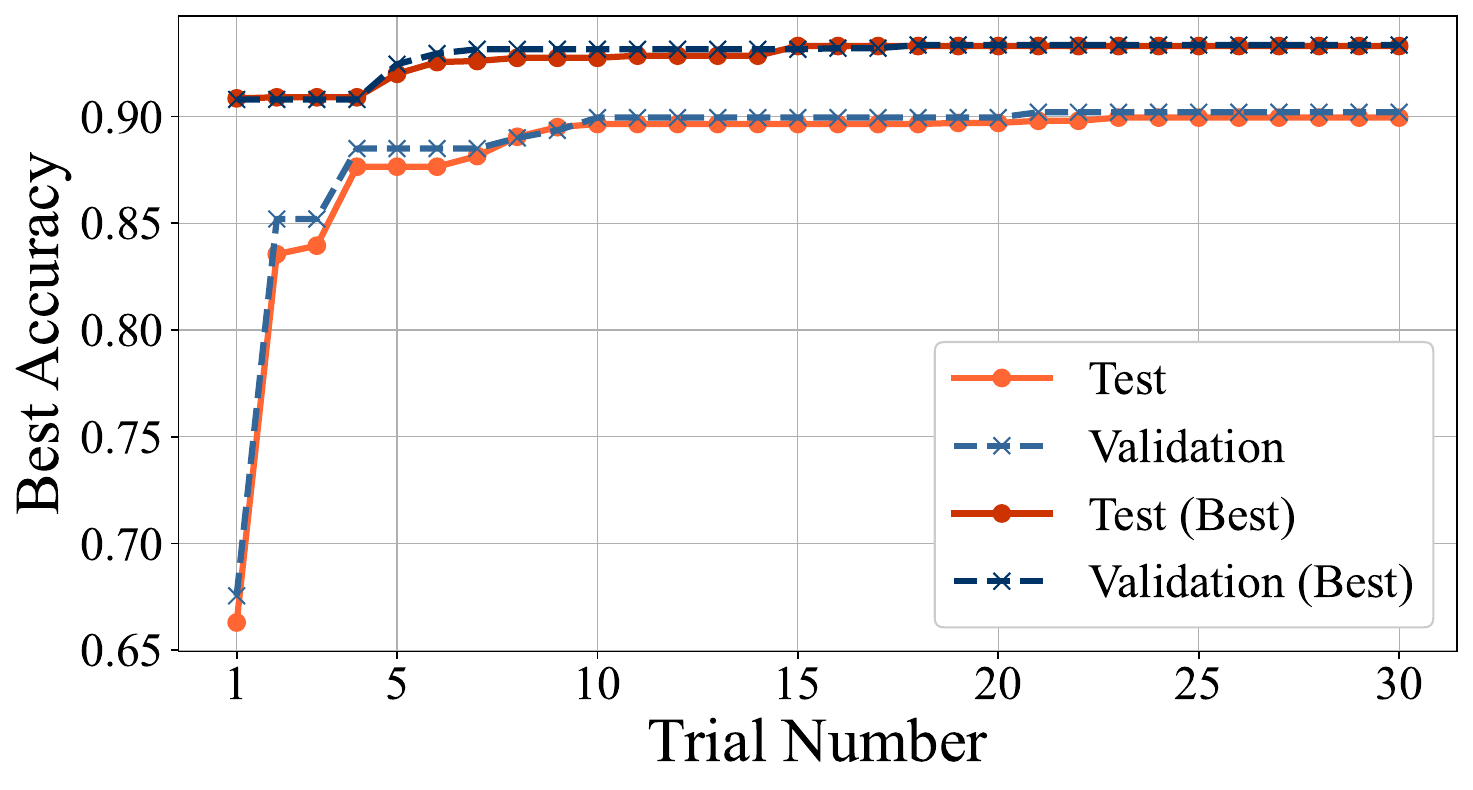}
    \caption{Trajectory of classification accuracy on the MNIST dataset using quantum feature maps generated by our agentic system. The curves shown in dark colors (red and blue) represent trials with high-performance initial ideas. The curves shown in lighter colors (orange and light blue) represent an example where low-performance initial ideas. The vertical axis of the figure represents the best validation accuracy up to that trial, defined as $\max(\mathrm{accuracy}(t), \mathrm{accuracy}(t-1), \dots)$, where $\mathrm{accuracy}(t)$ denotes the best validation accuracy in the $t$-th trial. The horizontal axis corresponds to the trial number.}
    \label{fig:learning_trajectory}
\end{figure}

Fig.~\ref{fig:learning_trajectory}, which shows two example trajectories, corresponding to the ones with high-performance and low-performance initial ideas, of the best validation/test accuracy across trials, demonstrates that the system progressively optimizes the quantum feature maps through iterative refinement.
The validation/test accuracy here is the one computed with the 2000 images passed to the agent for the validation and test, and not with the official test set of MNIST. 
We can also examine the feature maps generated throughout the trials to observe in what manner the system improves feature maps.

Let us first discuss the case where the system generated a high-performance initial idea.
In this particular case, the system initially generated a simple quantum circuit that first applies eight single-qubit rotations to each qubit, where each input dimension is embedded as the angle of a single-qubit rotation, and then applies $\prod_{i=0}^{9} \operatorname{CNOT}_{i,i+1}$, where $\operatorname{CNOT}_{i,j}$ denotes a controlled-NOT gate with control qubit $i$ and target qubit $j$.
This circuit, despite effectively being a quantum feature map consisting of single-qubit rotations only as the final CNOT gates cancels when taking the kernel values by Eq. \eqref{eq:kernel}, achieves over 90\% validation/test accuracy.
Note that this is not a surprising result; Ref.~\cite{huang2021power} has already shown that this type of entanglement-free feature map can achieve high-accuracy on a Fashion-MNIST dataset.
Subsequent trials from this initial idea introduced several refinements.
These include varying the type of rotation gates by layer index, adjusting scale parameters for rotation angles, and incorporating global features by aggregating all 80 input dimensions. 
A certain trial also added more advanced logic that independently tune scale parameters for each layer.
The embedding method also evolved from assigning a single data dimension to each gate to combining multiple data dimensions for a single rotation angle.
Ultimately, the system arrived at a complex circuit including many two-qubit gates and complicated, however linear, embedding of the input features to angles, which we show as Listing~\ref{lst:BestGeneratedCode} in Appendix~\ref{sec:featureMapCode}.
This feature map has the best performance over the whole in this work, achieving over $95\%$ in accuracy.

\begin{figure}[ht]
    \includegraphics[width=1.0\linewidth]{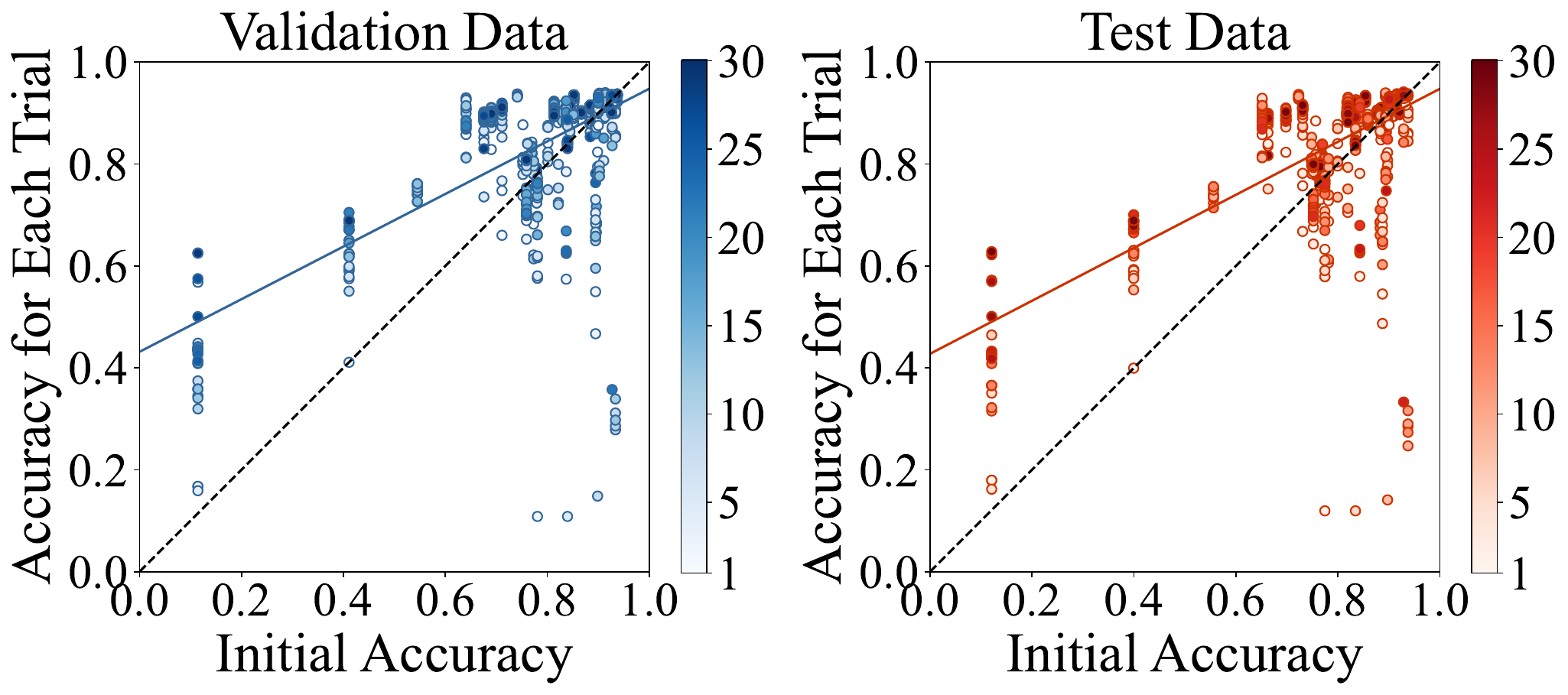}
    \caption{Trajectory of classification accuracy over the course of all 45 independent runs. The values on the Y-axis represent the accuracy obtained at each trial, rather than the best accuracy. Therefore, the highest value along the Y-axis corresponds to the final best accuracy. The color bars on the right side show the trial index. The number of trials is represented by the intensity of the plotted points, with darker colors indicating later trials.  The regression line represented by the solid line in the validation data is $y = 0.5164x + 0.4315$, and in the test data, it is $y = 0.5200x + 0.4276$. The black dotted line indicates the baseline; points above this line represent trials that exceeded the initial accuracy, while points below represent trials that fell short of it.}
    \label{fig:all_trajectories_trend}
\end{figure}

In contract, for cases where the system generated a low-performance initial idea, the performance tends to saturate at lower levels, as we will see later in Fig.~\ref{fig:all_trajectories_trend}.
An example of such trajectory is shown in Fig. ~\ref{fig:learning_trajectory} in a lighter color.
In this particular case, the two initial ideas were (1) a simple circuit that resembles the circuit generated in the high-performance gate but with significantly lower performance due to the different order of assigning feature values to angles, and (2) a complex circuit that involves controlled rotations with angles determined by multiple feature values but shows higher performance than (1).
We observe that the system pursued improvements in the idea (2) since it performed better than (1).
However, due to the complexity of the initial idea, the system fails to improve the feature maps to the level obtained in the high-performance case.

Finally, we show all trajectories of accuracy improvement across 45 runs in Fig.~\ref{fig:all_trajectories_trend}.
The same data but plotted against trial number can be found in Fig. ~\ref{fig:all_trajectories} of Appendix~\ref{sec:allTrajectoryHistory}.
The figure clearly shows the dependence of the obtained accuracies on the initial idea; the lower initial accuracy tends to result in lower accuracies in subsequent trials.
This result suggests that, if human professionals can provide good initial ideas, the system might be able to obtain feature maps that exceed the best result obtained in this work.
We leave developing such a system as an interesting future direction to explore.

\subsection{Generalizability of the best feature map}
Next, we evaluate the generalizability of the generated quantum feature maps on Fashion-MNIST and CIFAR-10 datasets.
The purpose of this evaluation is two-fold.
First, we wish to assure that the system has not designed a feature map that only works on MNIST dataset, which is used in the iterative improvement process. 
Knowledge about the MNIST dataset that is almost surely included in detail within LLMs can potentially lead them to design such a feature map.
Second objective is to compare the performance of the generated quantum feature maps among the ones that are widely used as baselines.
For classical feature maps to compare against, we select four kernels commonly used in SVMs: the linear kernel, the RBF kernel, the polynomial kernel, and the sigmoid kernel. 
For quantum feature maps, we selected the so-called ZZ feature map~\cite{havlivcek2019supervised}, the so-called NPQC feature map and  YZCX feature map~\cite{haug2021quantum}. 

Each method that we compare the generated feature map against involves multiple hyperparameters. 
Therefore, we employed Optuna~\cite{akiba2019optuna}, a hyperparameter optimization framework, to search for optimal parameter configurations based on the sampled training dataset, followed by evaluation on the corresponding test dataset. 
Since the optimization process involves inherent randomness, we conducted five independent optimization runs for each dataset. 

We report the results of each method obtained in this manner in Table ~\ref{tab:compare_results}. 
It shows the mean and standard deviation of classification accuracy obtained using the models configured with the best parameters from each run of the hyperparameter tuning.
From Table~\ref{tab:compare_results}, we can confirm that the generated quantum feature map achieved consistent classification accuracy across different datasets, indicating that they did not overfit to the MNIST dataset.
For all datasets, the generated feature map outperformed other widely used quantum feature maps in terms of classification accuracy. 
On the other hand, when compared to classical machine learning approaches, the generated maps outperformed the linear, polynomial and sigmoid kernels. 
However, they slightly underperformed compared to the RBF Kernel, which is the most commonly used and effective kernel in classical settings.

\subsection{Analysis beyond proof-of-concept}

Having established the basic performance and generalizability of our system, we now turn to a more in-depth analysis to demonstrate its capabilities beyond a mere proof-of-concept.

\subsubsection{Diversity of generated idea}

\input{tables/circuit_diversity}

Table~\ref{tab:diversityStats} shows the diversity evaluation results for each seed idea.
The mean distances are 0.3300 and 0.2244 for seed ideas 1 and 2, respectively.
For reference, we also compute the Frobenius distance between a linear kernel and a nonlinear RBF kernel under the same setting, obtaining a value of 0.1748. This comparison indicates that the generated quantum feature maps diverge meaningfully from the initial ideas, supporting the view that the system explores beyond a narrow local region of the initial design space.

Furthermore, the Frobenius distances between the best performing quantum feature map shown in Listing~\ref{lst:BestGeneratedCode} and the two seed ideas are 0.3091 and 0.1982, respectively. These values suggest that the system produced feature maps that are structurally distinct from each seed idea.

\subsubsection{Scalability of the best feature map}
\begin{figure}[h]
    \includegraphics[width=1.0\linewidth]{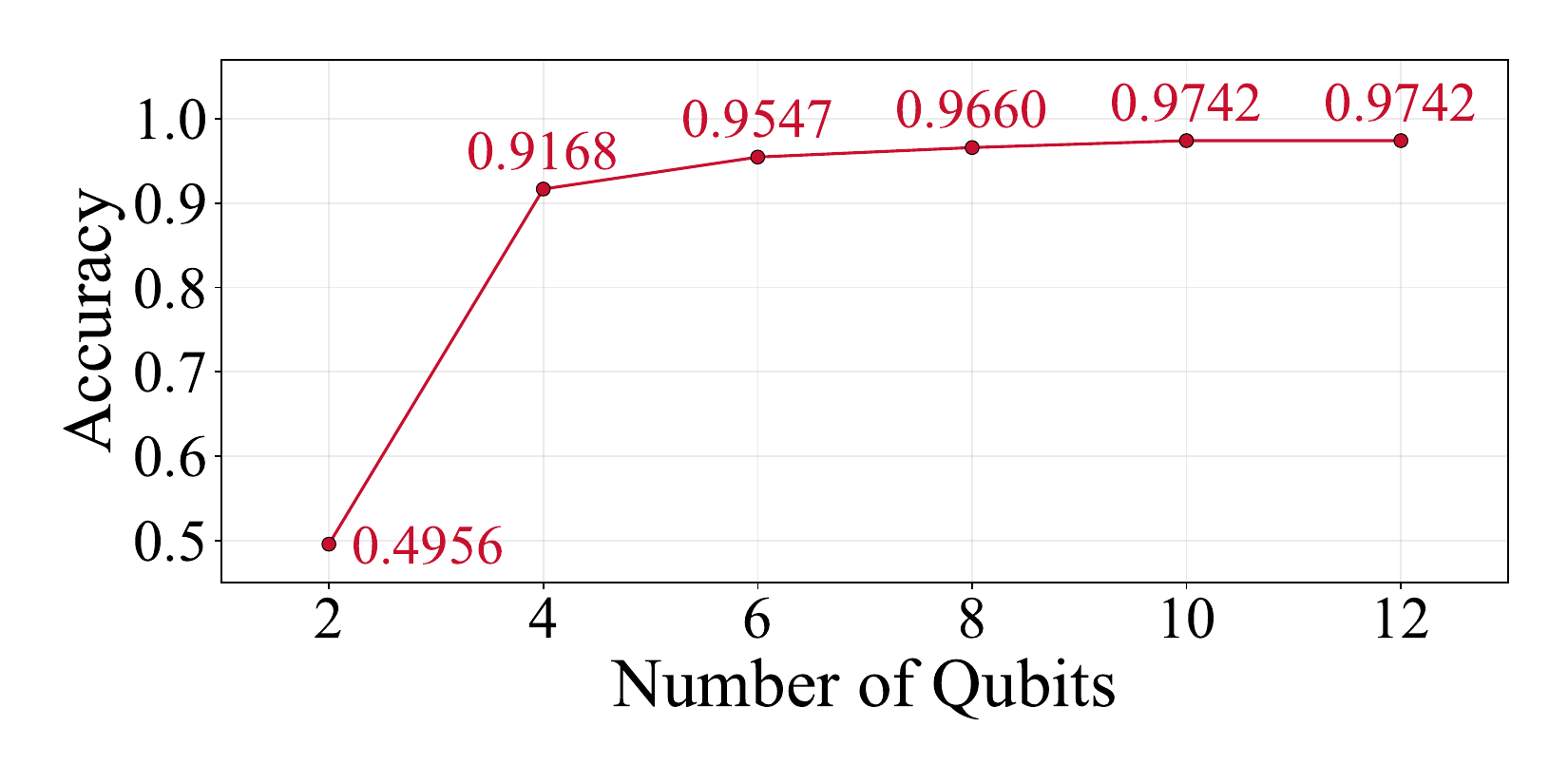}
    \caption{Accuracy on the full MNIST test set for generated circuits with different qubit counts, evaluated for systems ranging from two to twelve qubits in increments of two.}
    \label{fig:qubitScalability}
\end{figure}

Fig.~\ref{fig:qubitScalability} summarizes the MNIST test accuracy of the best feature map as the number of qubits increases.
We can see that the performance improves with the number of qubits and exceeds 90\% for four qubits or more.
The saturation of performance at ten qubits may arise because the generated circuit logic is originally optimized for the default qubit count. When applied to larger systems, the circuit can redundantly encode the same information across additional qubits, which does not provide further expressive power or accuracy gains.

\subsubsection{Comparison results with the QAS approaches}
\input{tables/compare_lei}

\input{tables/compare_zhang}

The comparison results with the QAS approaches are summarized in Tables~\ref{tab:compareLei} and \ref{tab:compareZhang}. Against Lei's study~\cite{lei2024neural}, our generated quantum feature map outperformed their results by approximately 5 to 6\% in classification accuracy under both evaluation conditions: (i) training on 300 samples with a disjoint 300-sample test set and (ii) an 80 to 20 train-test split. Similarly, in the comparison with Zhang's study~\cite{zhang2021neural}, our method achieved a 2.2\% accuracy gain over their reported results.

These results demonstrate that our system yields superior performance compared to existing QAS approaches. It is particularly notable that our method, which utilizes a general-purpose LLM for iterative refinement, outperforms prior works that rely on specialized neural networks tailored for architecture search. This indicates that a generalist LLM-driven approach can effectively navigate the design space and discover good quantum feature maps, surpassing the capabilities of dedicated, task-specific search algorithms.

\subsubsection{Ablation study results}
\begin{figure}[ht]
    \includegraphics[width=1.0\linewidth]{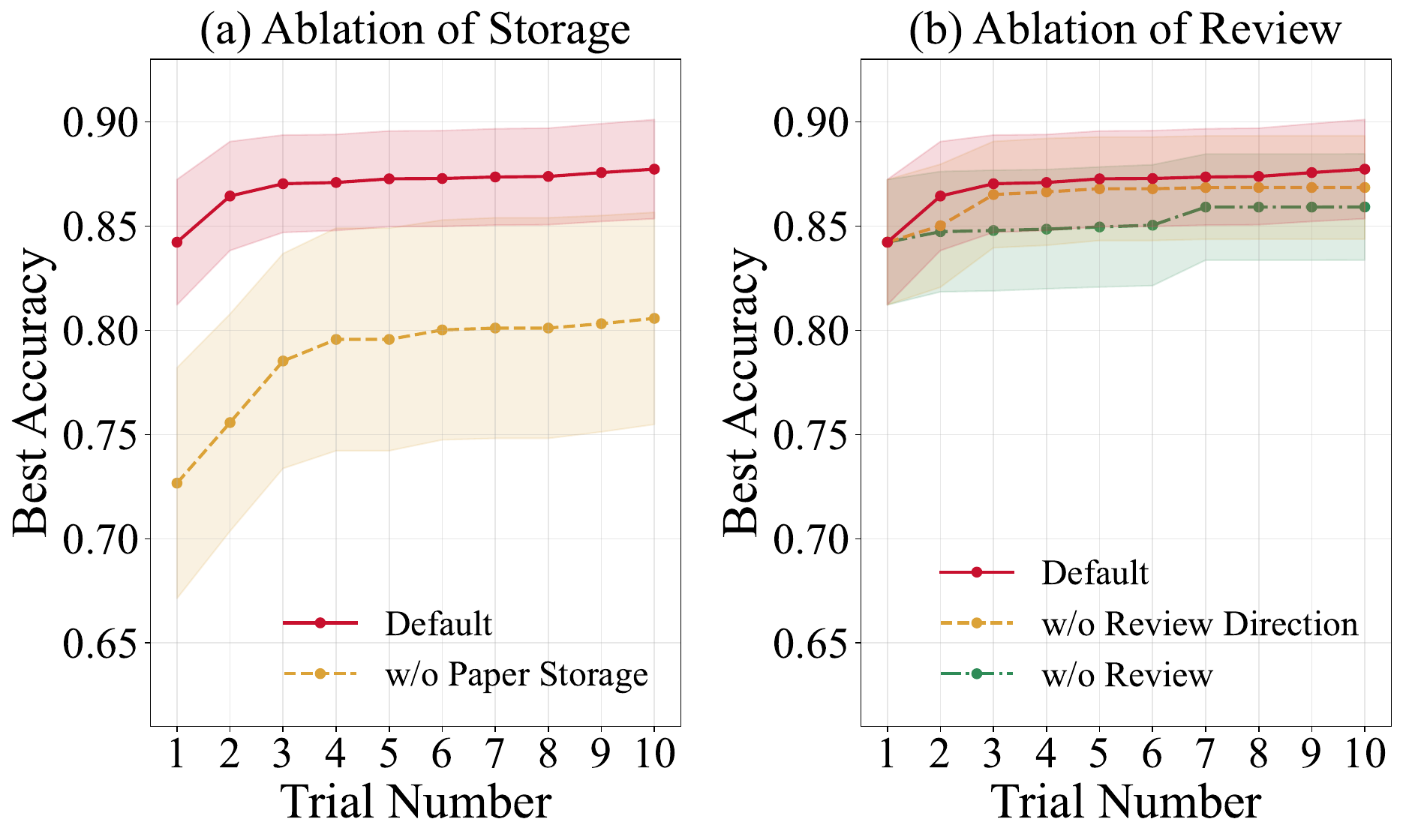}
    \caption{Results of the ablation study. Solid lines show the mean of the best accuracy across ten independent runs for each trial. Shaded bands denote the standard error of the mean. (a) Comparison between the default system and the version without the ``Storage'' component. (b) Comparison between the default system, the version without the direction determined by the $\mathrm{DiffMetric}$ values in ``Review'' component, and a version without the entire ``Review'' component.}
    \label{fig:ablation_study}
\end{figure}

Fig.~\ref{fig:ablation_study} shows the results of the ablation studies.
The solid lines report the mean of the best accuracy over ten independent runs at each trial. 
The shaded regions indicate the standard error of the mean.

The results of the first ablation study are shown in the left panel of Fig.~\ref{fig:ablation_study} (a). 
As we can see, removing access to recent academic papers causes an approximately 7.2\% decrease in the mean final accuracy.
The variation across trials shown by the shaded regions further shows that the default system (red) exhibits substantially better stability. More concretely, we find that the default system has about 46.8\% less variation in accuracy than the condition without paper storage (yellow).
These results indicate that the ``Storage'' component is important in generating effective quantum circuits.

The results of the second ablation study are shown in the right panel of Fig.~\ref{fig:ablation_study} (b). 
It shows that removing the direction determined by $\mathrm{DiffMetric}$ values in ``Review'' component (yellow) results in an approximately 0.9\% decrease in the mean final accuracy while the overlap of the shaded regions also shows that the difference might be due to coincidence.
Furthermore, removing the entire ``Review'' component (green) led to an approximately 1.8\% decrease in the mean final accuracy.
Although a positive contribution of the ``Review'' component is suggested, the observed variance prevents a definitive conclusion.
A more rigorous statistical validation is therefore required, which we leave for future research.

In summary, these ablation studies clarify the functional contributions of the individual components in the proposed system.
Removing the ``Storage'' component leads to a substantial degradation in both accuracy and stability, highlighting the importance of external knowledge retrieval.
In contrast, the ablation results for the ``Review'' component suggest a positive contribution to performance, although the observed variance prevents a definitive conclusion, supporting the overall architectural design.

\subsubsection{Design parameter variation results}
\begin{figure}[ht]
    \includegraphics[width=1.0\linewidth]{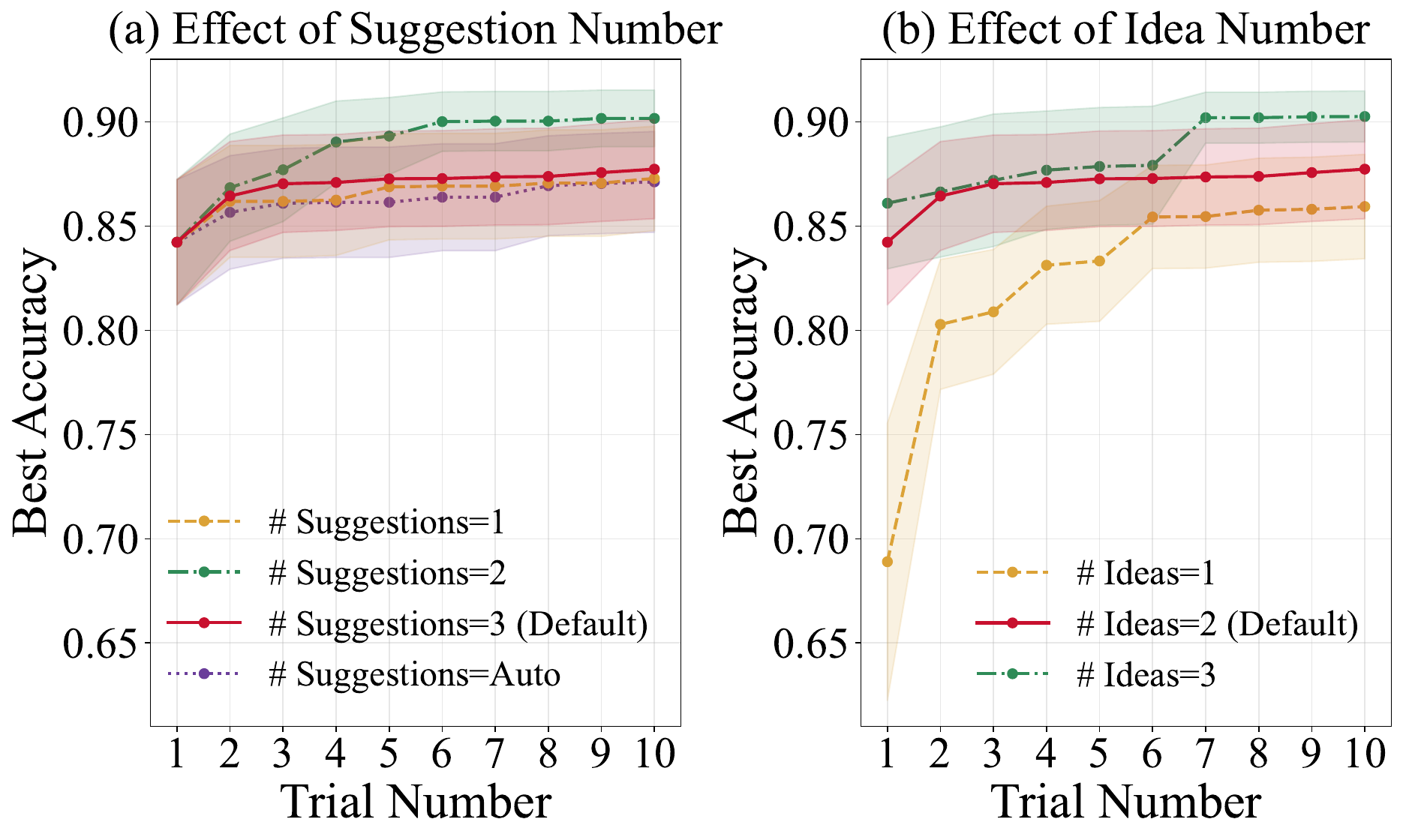}
    \caption{Results of the design parameter variation analysis. Solid lines show the mean of the best accuracy across ten independent runs for each trial. Shaded bands denote the standard error of the mean. (a) Comparison between the default system and the version with a reduced number of change suggestions in review step. (b) Comparison between the default system and the version with a number of ideas in the generation step.}
    \label{fig:params_variation}
\end{figure}

Fig.~\ref{fig:params_variation} shows the results of the parameter variation analysis.
The solid lines report the mean of the best accuracy over ten independent runs at each trial. 
The shaded regions indicate the standard error of the mean.

The results of the first parameter variation analysis are shown in  Fig.~\ref{fig:params_variation} (a). 
Among the explicitly specified settings, using two suggestions (green) yields the highest final mean accuracy and the most stable performance across trials, while both fewer and more suggestions lead to slightly degraded results.
When the number of suggestions is not explicitly specified in the prompt (purple), the system generates 4.22 suggestions per trial on average. Although this auto setting produces more suggestions, its final performance does not reach that of the explicitly specified settings.
These results indicate that increasing the number of suggestions can improve performance up to a certain point, beyond which additional suggestions do not yield further gains and may even degrade performance.
Suggestions produced by the ``Review'' component serve as supplementary material for the ``Generation'' component in the subsequent trial.
Since two ideas are generated per trial in this study, providing too many suggestions may hinder their effective incorporation, thereby limiting further gains. A systematic investigation of the balance between the number of generated ideas and suggestions is left for future work.

The results of the second parameter variation analysis are shown in Fig.~\ref{fig:params_variation} (b). 
It shows that increasing the number of generated ideas improves the mean performance and enhanced stability across trials.
This result suggests that it is better to use as large many ideas as possible.
However, the number of generated ideas directly affects both API usage and computational cost, which increased approximately linearly with this number. 
Determining an appropriate number of generated ideas therefore requires balancing performance gains against the associated costs, which we leave as future developments.
Also, note that generating more ideas per trial allowed the system to explore multiple directions, which improved trial to trial stability.
Since increasing the number of generated ideas effectively corresponds to increasing the sample size in trial, the observed improvement in stability with more ideas can be seen as a result of the law of large numbers.

In summary, the parameter variation analyses reveal that the performance of the proposed system is sensitive to key design choices in both the ``Review'' and ``Generation'' components.
The results indicate that an appropriate number of review suggestions is critical for achieving optimal performance, with overly few or excessive suggestions leading to degraded accuracy.
In addition, generating multiple ideas per trial improves both performance and stability, although this benefit must be balanced against the associated computational cost.

\section{Conclusion}
In this study, we proposed an agentic system that autonomously performs iterative refinement by using a LLM to generate ideas for quantum feature maps along with executable code, followed by reviewing the evaluation results.
Across multiple datasets, the resulting quantum feature maps achieve higher accuracy than several existing quantum feature maps, demonstrating the effectiveness of LLM-driven refinement guided by performance-based feedback.
While the proposed approach does not yet outperform strong classical machine learning models, it establishes a practical framework for automated discovery and improvement of quantum representations.

In future work, extending the system to support quantum machine learning models with trainable parameters—such as quantum circuit learning—and improving the logic of the iterative refinement process to enhance exploration efficiency may lead to the discovery of quantum feature maps that outperform classical machine learning models.

Beyond quantum feature map design, our agentic framework can potentially be extended to a broad class of variational quantum algorithms such as the variational quantum eigensolver (VQE)~\cite{Peruzzo2014} and the quantum approximate optimization algorithm (QAOA)~\cite{Farhi2014}. 
By enabling automatic circuit generation and refinement through empirical feedback, the system may help to discover more efficient or interpretable ansatz structures tailored to specific quantum tasks. 
Furthermore, adapting the system to generate and optimize circuits for quantum error correction, or to assist in the construction of novel quantum algorithms, presents a promising direction for future research at the intersection of quantum computing and autonomous scientific discovery.

\section{Code and data availability}
\label{sec:sourceCode}
The source code of the agentic system and the results reported in this study are publicly available at \url{https://github.com/Qyusu/astronaut}.

\begin{acknowledgments}
This work is supported by MITOU Target Program, organized by Information-technology Promotion Agency (IPA).
This work is also supported by MEXT Quantum Leap Flagship Program (MEXT Q-LEAP) Grant Nos. JPMXS0120319794 and JPMXS0118067394, and JST COI-NEXT Grant No. JPMJPF2014.
K.M. is supported by JST FOREST Grant No. JPMJFR232Z and JSPS KAKENHI Grant No. 23H03819.
\end{acknowledgments}

\bibliography{references}

\clearpage
\input{appendix}

\end{document}

%% file: code_style.tex
\usepackage{xcolor}
\usepackage{listings}

\lstdefinestyle{promptFormat}{
    language=Python,
    inputencoding=utf8,
    basicstyle=\ttfamily\small,            
    keywordstyle=\color{red}\bfseries,     
    commentstyle=\color{green!50!black},   
    stringstyle=\color{blue},              
    backgroundcolor=\color{gray!10},       
    frame=single,                           
    showstringspaces=false,                 
    breaklines=true,                         
    numbers=left,                            
    numberstyle=\tiny\color{gray},           
    numbersep=5pt,                           
    xleftmargin=10pt,                        
    literate={\{max\_trial\_num\}}{{\textcolor{red}{\{max\_trial\_num\}}}}{14}
             {\{current\_trial\}}{{\textcolor{red}{\{current\_trial\}}}}{15}
             {\{device\_n\_qubit\}}{{\textcolor{red}{\{device\_n\_qubit\}}}}{16}
             {\{idea\_num\}}{{\textcolor{red}{\{idea\_num\}}}}{10}
             {\{previous\_trial\}}{{\textcolor{red}{\{previous\_trial\}}}}{16}
             {\{review\_comment\}}{{\textcolor{red}{\{review\_comment\}}}}{16}
             {\{few\_shot\_examples\}}{{\textcolor{red}{\{few\_shot\_examples\}}}}{19}
             {\{current\_round\}}{{\textcolor{red}{\{current\_round\}}}}{15}
             {\{max\_scoring\_round\}}{{\textcolor{red}{\{max\_scoring\_round\}}}}{19}
             {\{max\_reflection\_round\}}{{\textcolor{red}{\{max\_reflection\_round\}}}}{22}
             {\{previous\_idea\}}{{\textcolor{red}{\{previous\_idea\}}}}{15}
             {\{previous\_score\}}{{\textcolor{red}{\{previous\_score\}}}}{16}
             {\{related\_work\}}{{\textcolor{red}{\{related\_work\}}}}{14}
             {\{max\_summary\_words\}}{{\textcolor{red}{\{max\_summary\_words\}}}}{19}
             {\{raw\_content\}}{{\textcolor{red}{\{raw\_content\}}}}{13}
             {\{code\}}{{\textcolor{red}{\{code\}}}}{6}
             {\{idea\}}{{\textcolor{red}{\{idea\}}}}{6}
             {\{pennylane\_operations\}}{{\textcolor{red}{\{pennylane\_operations\}}}}{22}
             {\{methods\}}{{\textcolor{red}{\{methods\}}}}{9}
             {\{references\}}{{\textcolor{red}{\{references\}}}}{12}
             {\{error\_messages\_string\}}{{\textcolor{red}{\{error\_messages\_string\}}}}{23}
             {\{warning\_messages\_string\}}{{\textcolor{red}{\{warning\_messages\_string\}}}}{25}
             {\{max\_suggestion\_num\}}{{\textcolor{red}{\{max\_suggestion\_num\}}}}{20}
             {\{last\_trial\_num\}}{{\textcolor{red}{\{last\_trial\_num\}}}}{16}
             {\{last\_trial\_results\}}{{\textcolor{red}{\{last\_trial\_results\}}}}{20}
             {\{quantum\_gate\_list\}}{{\textcolor{red}{\{quantum\_gate\_list\}}}}{19}
             {α}{{$\alpha$}}1
             {β}{{$\beta$}}1
             {γ}{{$\gamma$}}1
             {∑}{{$\sum$}}1
             {θ}{{$\theta$}}1
             {φ}{{$\phi$}}1
             {Φ}{{$\Phi$}}1
             {ψ}{{$\psi$}}1
             {ᵧ}{{$^\gamma$}}1
             {ₓ}{{$_x$}}1
             {ᵢ}{{$^i$}}1
             {ⱼ}{{$^j$}}1
             {₀}{{$_0$}}1
             {₁}{{$_1$}}1
             {₂}{{$_2$}}1
             {₃}{{$_3$}}1
             {₄}{{$_4$}}1
             {₅}{{$_5$}}1
             {₆}{{$_6$}}1
             {₇}{{$_7$}}1
             {₈}{{$_8$}}1
             {ⱼ}{{$_j$}}1
             {π}{{$\pi$}}1
             {…}{{$\ldots$}}1
             {×}{{$\times$}}1
             {²}{{$^2$}}1
             {√}{{$\sqrt{}$}}1
             {≤}{{$\le$}}1
             {∝}{{$\propto$}}1
             {“}{{"}}1
             {”}{{"}}1
             {‘}{{'}}1
             {’}{{'}}1
             {→}{{$\rightarrow$}}1
             {⟩}{{$\rangle$}}1
             {⟨}{{$\langle$}}1
             {≥}{{$\ge$}}1
}

\lstdefinestyle{codeFormat}{
    language=Python,
    inputencoding=utf8,
    basicstyle=\ttfamily\small,            
    keywordstyle=\color{red}\bfseries,     
    commentstyle=\color{green!50!black},   
    stringstyle=\color{blue},              
    backgroundcolor=\color{gray!10},       
    frame=single,                           
    showstringspaces=false,                 
    breaklines=true,                         
    numbers=left,                            
    numberstyle=\tiny\color{gray},           
    numbersep=5pt,                           
    xleftmargin=10pt,                        
    literate={π}{{$\pi$}}1
             {₁}{{$_1$}}1
             {λ}{{$\lambda$}}1
             {ₗ}{{$_\ell$}}1
             {γ}{{$\gamma$}}1
             {Σ}{{$\sum$}}1
             {∇}{{$\nabla$}}1
             {Δ}{{$\Delta$}}1
             {→}{{$\rightarrow$}}1
}

\lstdefinestyle{txtFormat}{
    language=,                               
    inputencoding=utf8,
    basicstyle=\ttfamily\small,              
    keywordstyle=,                           
    commentstyle=,                           
    stringstyle=,                            
    backgroundcolor=\color{gray!10},         
    frame=single,                            
    showstringspaces=false,                  
    breaklines=true,                         
    numbersep=5pt,                           
    xleftmargin=10pt,                        
    literate={π}{{$\pi$}}1
             {σ}{{$\sigma$}}1
             {ρ}{{$\rho$}}1
             {γ}{{$\gamma$}}1
             {ℓ}{{$\ell$}}1
             {α}{{$\alpha$}}1
             {Σ}{{$\Sigma$}}1
             {Δ}{{$\Delta$}}1
             {ξ}{{$\xi$}}1
             {ε}{{$\varepsilon$}}1
             {φ}{{$\varphi$}}1
             {τ}{{$\tau$}}1
             {‖}{{$\lVert$}}1
             {→}{{$\rightarrow$}}1
             {≤}{{$\leq$}}1
             {≥}{{$\geq$}}1
             {∈}{{$\in$}}1
             {≈}{{$\approx$}}1
             {₁}{{$_1$}}1
             {λ}{{$\lambda$}}1
             {ₗ}{{$_\ell$}}1
             {∇}{{$\nabla$}}1
             {→}{{$\rightarrow$}}1
             {±}{{$\pm$}}1
             {⌈}{{$\lceil$}}1
             {⌉}{{$\rceil$}}1
             {·}{{$\cdot$}}1
             {・}{{$\cdot$}}1
             {≪}{{$\ll$}}1
             {∏}{{$\prod$}}1
             {θ}{{$\theta$}}1
             {β}{{$\beta$}}1
             {ŵ}{{$\hat{w}$}}1
             {↔}{{$\leftrightarrow$}}1
             {⊗}{{$\otimes$}}1
             {…}{{$\dots$}}1
             {κ}{{$\kappa$}}1
             {×}{{$\times$}}1
             {≠}{{$\neq$}}1
             {μ}{{$\mu$}}1
             {↦}{{$\mapsto$}}1
             {←}{{$\gets$}}1
             {ϵ}{{$\epsilon$}}1
             {δ}{{$\delta$}}1
}

%% file: tables/feature_map_performance.tex
\begin{table*}
\caption{Generated quantum feature map performance on different datasets.}
\label{tab:compare_results}
\setlength{\tabcolsep}{4pt}
\begin{ruledtabular}
\begin{tabular}{llrrr}
Type & Method & \multicolumn{1}{c}{MNIST} & \multicolumn{1}{c}{Fashion-MNIST} & \multicolumn{1}{c}{CIFAR-10} \\
\hline
Classical & Linear kernel
  & 0.9385 $\pm$ 0.0002 & 0.8437 $\pm$ 0.0009 & 0.4087 $\pm$ 0.0011 \\
 & Polynomial kernel
  & 0.9667 $\pm$ 0.0058 & 0.8702 $\pm$ 0.0030 & 0.5375 $\pm$ 0.0014 \\
 & Sigmoid kernel
  & 0.9343 $\pm$ 0.0002 & 0.8189 $\pm$ 0.0120 & 0.4079 $\pm$ 0.0006 \\
 & RBF kernel
  & \textbf{0.9765} $\pm$ 0.0005
  & \textbf{0.8864} $\pm$ 0.0014
  & \textbf{0.5669} $\pm$ 0.0085 \\
Quantum & ZZ feature map \cite{havlivcek2019supervised}
  & 0.9255 $\pm$ 0.0009 & 0.8252 $\pm$ 0.0023 & 0.3907 $\pm$ 0.0016 \\
 & NPQC feature map \cite{haug2021quantum}
  & 0.9644 $\pm$ 0.0028 & 0.8749 $\pm$ 0.0026 & 0.4903 $\pm$ 0.0188 \\
 & YZCX feature map \cite{haug2021quantum}
  & 0.9727 $\pm$ 0.0006 & 0.8778 $\pm$ 0.0049 & 0.4753 $\pm$ 0.0341 \\
 & Generated (ours)
  & \textbf{0.9731} $\pm$ 0.0008
  & \textbf{0.8835} $\pm$ 0.0021
  & \textbf{0.5290} $\pm$ 0.0030 \\
\end{tabular}
\end{ruledtabular}
\end{table*}

%% file: tables/circuit_diversity.tex
\begin{table}
\caption{Normalized Frobenius distances between kernel matrices of seed feature maps from Trial 1 and those from subsequent trials.}
\label{tab:diversityStats}
\begin{ruledtabular}
\begin{tabular}{lcccc}
Seed & Mean & Std & Min & Max \\
\hline
Seed 1 & 0.3300 & 0.0398 & 0.2791 & 0.5494 \\
Seed 2 & 0.2244 & 0.0446 & 0.1494 & 0.4607 \\
\end{tabular}
\end{ruledtabular}
\end{table}

%% file: tables/compare_lei.tex
\begin{table}[t]
\caption{Comparison with Lei's study using a subset of the MNIST dataset.}
\label{tab:compareLei}
\begin{ruledtabular}
\begin{tabular}{lr}
Method & Accuracy \\
\hline
Lei \textit{et al.} \cite{lei2024neural} & 0.9027 $\pm$ 0.0053 \\
Ours, setting (i) & \textbf{0.9533} \\
Ours, setting (ii) & \textbf{0.9667} \\
\end{tabular}
\end{ruledtabular}
\end{table}

%% file: tables/compare_zhang.tex
\begin{table}[t]
\caption{Comparison with Zhang's study using a subset of the Fashion-MNIST dataset.}
\label{tab:compareZhang}
\begin{ruledtabular}
\begin{tabular}{lr}
Method & Accuracy \\
\hline
Zhang \textit{et al.} \cite{zhang2021neural} & 0.924 \\
Ours & \textbf{0.946} \\
\end{tabular}
\end{ruledtabular}
\end{table}

%% file: appendix.tex
\appendix
\onecolumngrid

\section{Generated Feature Map Code}\label{sec:featureMapCode}

In the developed agent system, an LLM automatically generated executable Python code for each trial.
The ``Generation'' component of idea produces the text show in Listing~\ref{lst:featureMapIdea} and ~\ref{lst:featureMapFormula} when the best performing quantum feature map is identified. This text is then used as system design document in the ``Generation'' component of code.
The resulting code is presented in Listing~\ref{lst:BestGeneratedCode}, and the corresponding quantum circuit diagram is shown in Figure~\ref{fig:best_circuit}. 

\begin{lstlisting}[style=txtFormat, caption={The feature map idea generated by our agentic system}, label=lst:featureMapIdea]
# feature_map_name 
Adaptive Single-Axis with Mid-Range CRot and ISWAP Fusion Feature Map

# summary
This design employs a streamlined single-axis local encoding using RY(π·x) rotations for minimal circuit depth. It incorporates adaptive mid-range entanglement with immediate neighbor coupling via CRX gates, next-nearest neighbor coupling via CRY gates, and a calibrated mid-range CRot stage using an adaptive scaling factor λₗ derived from data-specific noise profiles. An ISWAP layer, applied pairwise between qubits with maximal separation (specifically, qubit j with qubit (j+5) mod 10), is integrated to capture non-nearest interactions, while global aggregation is achieved via a MultiRZ gate with optimized non-uniform weights based on advanced metrics. Comprehensive error mitigation techniques (e.g., zero-noise extrapolation, measurement twirling, optimized circuit scheduling, and pulse-efficient transpilation) further bolster robustness.

# explanation
In this approach, the 80-dimensional PCA-reduced input is partitioned into five layers of 16 features. Within each layer, the first 10 features are encoded on a 10-qubit register using fixed RY(π·x) rotations, ensuring efficient local encoding with low circuit depth. Immediate neighbor entanglement is realized via CRX gates, where the rotation angles are calculated as π times a linear combination (a uniform weighted pairwise average with an added contrast term) of adjacent features, whereas next-nearest neighbor coupling is performed using CRY gates based on an equal-weight triple average. To capture medium-range correlations, a CRot-based entanglement stage couples qubits separated by three positions, with an adaptive, noise-calibrated scaling factor λₗ. Crucially, an ISWAP layer is incorporated where gates are applied pairwise-specifically connecting qubit j with qubit (j+5) mod 10-to effectively integrate non-nearest neighbor correlations. Finally, a global MultiRZ gate aggregates features from all layers using optimized non-uniform weights derived from variance and separability analyses, with integrated error mitigation strategies maintaining performance despite noise. Moreover, the architecture is amenable to further optimization via pulse-efficient transpilation techniques, which can reduce circuit duration and mitigate dephasing errors on near-term quantum processors.

# key_sentences
- The input is partitioned into 5 layers of 16 features, with each layer's first 10 features encoded using RY(π·x) rotations on a 10-qubit register, ensuring robust local encoding with minimal circuit depth.
- Immediate neighbor entanglement is uniformly implemented with CRX gates, where the rotation angle is computed as π times a uniform weighted average of a pair of entanglement features plus a low fixed contrast term, while next-nearest neighbor coupling is executed with CRY gates using an equal-weight triple average.
- A mid-range entanglement stage is incorporated via CRot gates to couple qubits separated by three positions, with an adaptive, noise-calibrated scaling factor λₗ that balances enhanced correlation capture with noise suppression.
- An ISWAP layer is introduced, applying gates pairwise between qubit j and qubit (j+5) mod 10, to capture non-nearest neighbor correlations and enrich the entangled state.
- Global aggregation is performed via a MultiRZ gate with optimized non-uniform weights (Δₗ) derived from variance and separability analyses, with advanced error mitigation techniques ensuring robust performance in noisy environments.
- The architecture is further conducive to optimization via pulse-efficient transpilation, aligning with recent advances in reducing circuit duration and mitigating decoherence.
\end{lstlisting}

\Needspace{22\baselineskip}
\begin{minipage}{\linewidth}
    \listingcaption{Best Feature Map Formula}
    \label{lst:featureMapFormula}
    \input{ideas/best_feature_map_formula}
\end{minipage}

\input{circuits/best_circuit}

\begin{lstlisting}[style=codeFormat, caption={Generated feature map code by our agentic system}, label=lst:BestGeneratedCode]
import numpy as np
import pennylane as qml
from qxmt.constants import PENNYLANE_PLATFORM
from qxmt.feature_maps import BaseFeatureMap

# new imports can be added below this line if needed.

class AdaptiveSingleAxisWithMidRangeCRotAndISWAPFusionFeatureMap(BaseFeatureMap):
    """
    Adaptive Single-Axis with Mid-Range CRot and ISWAP Fusion Feature Map.
    
    This feature map partitions the 80-dimensional PCA-reduced input into 5 layers (each with 16 features).
    For each layer l (l = 1,...,5):
      
    - Local Encoding:
         The first 10 features are encoded on a 10-qubit register using fixed RY(π * x) rotations,
         ensuring efficient local encoding.
      
    - Stage 1 (Immediate Neighbor Entanglement):
         For each qubit j, two designated entanglement features are selected from the block starting at index 16*l + 10:
           x_a = x[16*l + 10 + (j mod 6)]
           x_b = x[16*l + 10 + ((j+1) mod 6)]
         The rotation angle is computed as:
           angle = π * (0.5*x_a + 0.5*x_b + 0.1*(x_a - x_b))
         A CRX gate is applied between qubit j and (j+1) mod n_qubits.
      
    - Stage 2 (Next-Nearest Neighbor Entanglement):
         For each qubit j, three features are selected at indices (j mod 6), ((j+2) mod 6), and ((j+4) mod 6)
         from the same block. Their average is used to compute the rotation angle (π times the average),
         and a CRY gate is applied between qubit j and (j+2) mod n_qubits.
      
    - Stage 3 (Mid-Range Entanglement):
         For each qubit j, the average of the same two features as in Stage 1 (x_a and x_b) is computed,
         scaled by an adaptive factor λₗ (provided via lambda_factors), yielding an angle:
           angle = π * λₗ * (0.5*x_a + 0.5*x_b)
         A CRot gate (with theta=0.0 and omega=0.0) is applied between qubit j and (j+3) mod n_qubits.
      
    - Stage 4 (ISWAP Fusion):
         To integrate non-nearest neighbor correlations, an ISWAP-like interaction is applied using a parameterized
         IsingXY gate. For j = 0,...,n_qubits/2 - 1 (to avoid duplication), the gate is applied on qubits j and j + n_qubits/2
         with rotation angle π * γ, where γ is a scaling factor.
      
    - Global Entanglement:
         A MultiRZ gate aggregates features from all layers with a rotation angle computed as:
           global_angle = π * Σₗ [ Δₗ * x[16*l+10] ]
         where Δₗ are fixed non-uniform weights.
      
    Note: The input x is expected to have shape (80,).
    
    Parameters:
      n_qubits (int): Number of qubits (ideally 10).
      lambda_factors (list): A list of 5 adaptive scaling factors for Stage 3. Default is [0.3, 0.3, 0.3, 0.3, 0.3].
      delta_weights (list): A list of 5 weights for global entanglement. Default is [0.15, 0.25, 0.35, 0.15, 0.10].
      gamma (float): Scaling factor for the ISWAP-like fusion interaction. Default is 0.5.
    """
    def __init__(self, n_qubits: int, lambda_factors: list = None, delta_weights: list = None, gamma: float = 0.5) -> None:
        super().__init__(PENNYLANE_PLATFORM, n_qubits)
        self.n_qubits = n_qubits
        
        # Adaptive scaling factors for mid-range entanglement (Stage 3)
        if lambda_factors is None:
            self.lambda_factors = [0.3, 0.3, 0.3, 0.3, 0.3]
        else:
            self.lambda_factors = lambda_factors
        
        # Global entanglement weights
        if delta_weights is None:
            self.delta_weights = [0.15, 0.25, 0.35, 0.15, 0.10]
        else:
            self.delta_weights = delta_weights
        
        # Scaling factor for the ISWAP fusion stage (implemented via IsingXY gate)
        self.gamma = gamma
        
    def feature_map(self, x: np.ndarray) -> None:
        expected_length = 5 * 16
        if len(x) != expected_length:
            raise ValueError(f"Input data dimension must be {expected_length}, but got {len(x)}")
        
        # Process each of the 5 layers
        for l in range(5):
            base = 16 * l
            
            # Local Encoding: Apply RY rotations for qubits 0 through 9
            for j in range(self.n_qubits):
                angle_ry = np.pi * x[base + j]
                qml.RY(phi=angle_ry, wires=j)
            
            # Stage 1: Immediate Neighbor Entanglement using CRX gates
            for j in range(self.n_qubits):
                idx_a = base + 10 + (j % 6)
                idx_b = base + 10 + ((j + 1) % 6)
                x_a = x[idx_a]
                x_b = x[idx_b]
                angle_immediate = np.pi * (0.5 * x_a + 0.5 * x_b + 0.1 * (x_a - x_b))
                qml.CRX(phi=angle_immediate, wires=[j, (j + 1) % self.n_qubits])
            
            # Stage 2: Next-Nearest Neighbor Entanglement using CRY gates
            for j in range(self.n_qubits):
                idx1 = base + 10 + (j % 6)
                idx2 = base + 10 + ((j + 2) % 6)
                idx3 = base + 10 + ((j + 4) % 6)
                avg_triple = (x[idx1] + x[idx2] + x[idx3]) / 3.0
                angle_cry = np.pi * avg_triple
                qml.CRY(phi=angle_cry, wires=[j, (j + 2) % self.n_qubits])
            
            # Stage 3: Mid-Range Entanglement using CRot gates
            for j in range(self.n_qubits):
                idx_a = base + 10 + (j % 6)
                idx_b = base + 10 + ((j + 1) % 6)
                pair_avg = 0.5 * (x[idx_a] + x[idx_b])
                angle_cret = np.pi * self.lambda_factors[l] * pair_avg
                qml.CRot(phi=angle_cret, theta=0.0, omega=0.0, wires=[j, (j + 3) % self.n_qubits])
            
            # Stage 4: ISWAP Fusion Layer using a parameterized IsingXY gate
            # Apply the gate on pairs to avoid duplication. For 10 qubits, apply on pairs (0,5), (1,6), ..., (4,9).
            for j in range(self.n_qubits // 2):
                qml.IsingXY(phi=np.pi * self.gamma, wires=[j, j + self.n_qubits // 2])
        
        # Global Entanglement: Aggregate inter-layer features via a MultiRZ gate
        global_sum = 0.0
        for l in range(5):
            global_sum += self.delta_weights[l] * x[16 * l + 10]
        global_angle = np.pi * global_sum
        qml.MultiRZ(theta=global_angle, wires=list(range(self.n_qubits)))
\end{lstlisting}

\section{All Trajectories of 45 Independent Runs by Our Agentic System}\label{sec:allTrajectoryHistory}
\begin{figure}[ht]
    \includegraphics[width=0.95\linewidth]{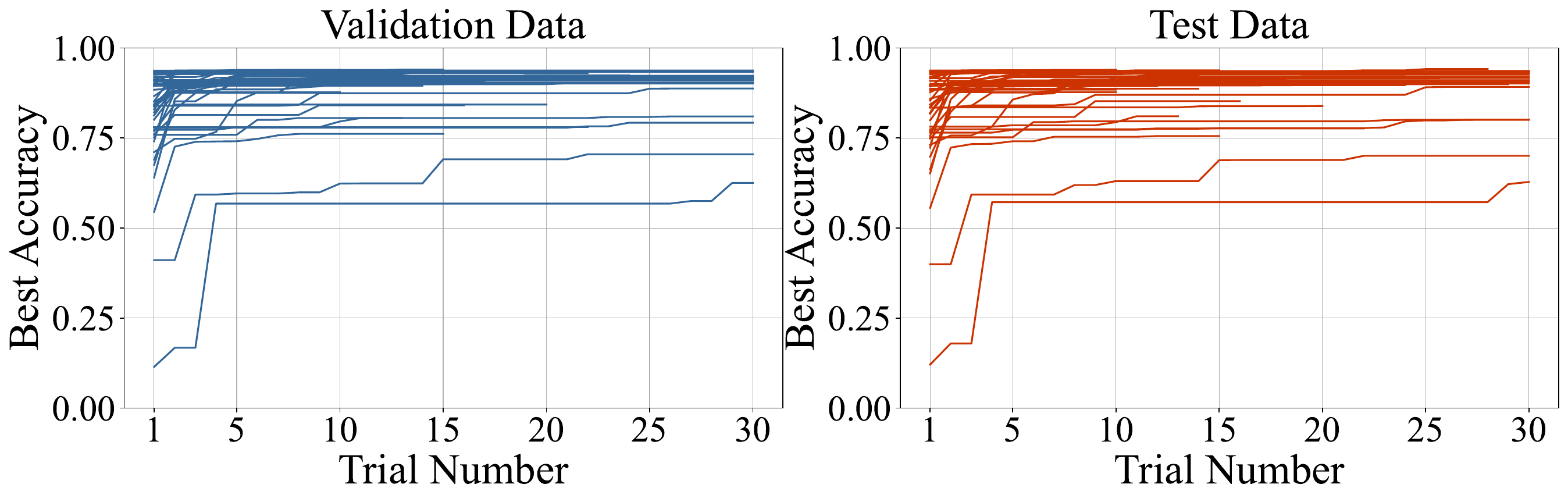}
    \caption{The trajectories of accuracy achieved by the Agentic System are plotted for all 45 independent runs. The left side shows the results on the validation data, while the right side shows the results on the test data.}
    \label{fig:all_trajectories}
\end{figure}

All 45 independent runs were conducted using the same architecture and prompt.
For each run, the trajectory of the best accuracy at each trial is plotted in Figure~\ref{fig:all_trajectories}. The method for calculating best accuracy, which is shown on the Y-axis, is the same as in Figure~\ref{fig:learning_trajectory}. The system is evaluated using validation and test dataset that described in Section~\ref{sec:datasetPreprocessing}.

\section{Supplementary Evaluation on Alternative LLMs}\label{sec:otherModels}
In this study, we employed a LLM developed by OpenAI within our system.
However, a wide range of LLMs is available regardless of their form of release, including open-source models and closed models accessible via APIs provided by developers.
Furthermore, the development of LLMs specialized for specific domains or tasks is actively progressing.
To the best of our knowledge, there is currently no publicly available LLM specifically designed for the quantum domain.
Therefore, for comparative purposes, we selected two general-purpose LLMs that are conceptually aligned with the OpenAI model: Gemini developed by Google~\cite{google2023gemini} and Claude developed by Anthropic~\cite{anthropic2024claude3}.
It should be noted that the prompts used in this study were optimized for the OpenAI model and have not been sufficiently tuned for Gemini or Claude.
Accordingly, the results presented in this section are intended to demonstrate that the proposed agentic system can function across different LLMs, rather than to provide a quantitative comparison of performance among LLMs.
The hyperparameters used in the evaluations are the same as those described in Appendix \ref{tab:hyperparams}, except that \texttt{max\_iter} was reduced to 15, and \texttt{idea\_num} was increased to 3.

\begin{figure}[ht]
    \includegraphics[width=0.8\linewidth]{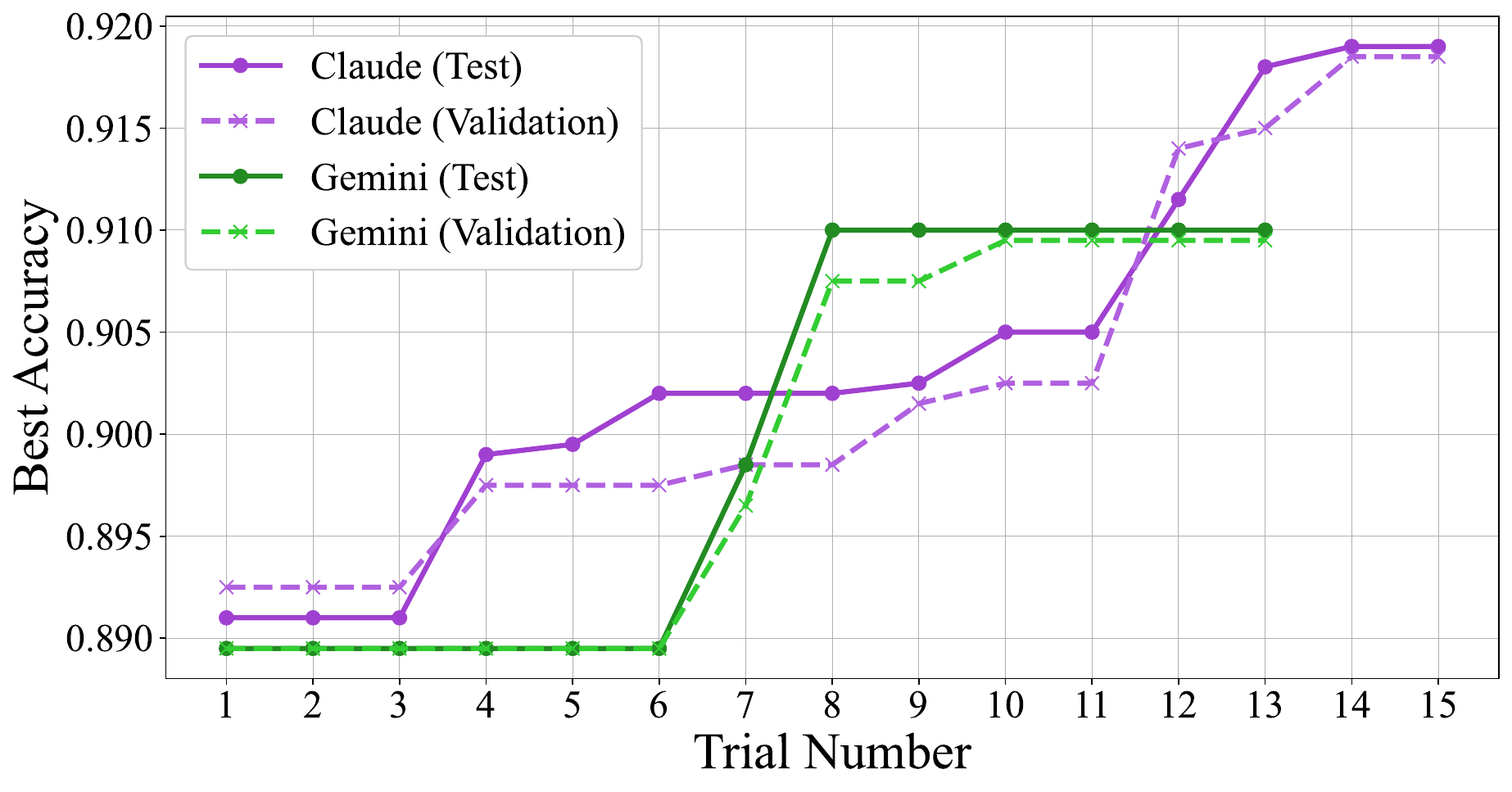}
    \caption{Trajectory of classification accuracy on the MNIST dataset. 
The green line represents using the Gemini model as the LLM, while the purple line represents using the Claude model.}
    \label{tab:compare_other_model}
\end{figure}

\subsection{Gemini by Google}
The Gemini series includes multiple models designed for different types of tasks. To maintain consistency with the model-task correspondence used for OpenAI’s models, we assigned specific Gemini models to each task category. In particular, ``gemini-2.0-pro-exp-02-05'' was used for review, idea-generation, idea-reflection, and code-generation tasks; ``gemini-2.0-flash-001'' was used for scoring and validation; and ``gemini-2.0-flash-lite-001'' was used for summary tasks.

We confirmed that the proposed agentic system is also capable of generating and iteratively improving quantum feature maps when using the Gemini models (Figure \ref{tab:compare_other_model}, green line). In the case of the Gemini model, the termination of the improvement process was determined by the LLM during the ``Review'' component following Trial 13. It was also observed that, when using Gemini, the system tends to produce relatively simple quantum feature maps and implementations (Listing \ref{lst:GeminiGeneratedCode}). In addition, the average processing time per trial was found to be 14 minutes, which is faster compared to other models (OpenAI model is 32 minutes, Claude model is 44 minute). Although all models were accessed via APIs and thus subject to network conditions and latency, the inference response time— even when using reasoning models— was fast overall, even when accounting for such factors.

\begin{lstlisting}[language=Python, inputencoding=utf8, style=codeFormat, caption={Generated feature map code by Gemini model}, label=lst:GeminiGeneratedCode]
import numpy as np
import pennylane as qml
from qxmt.constants import PENNYLANE_PLATFORM
from qxmt.feature_maps import BaseFeatureMap

# above are the default imports. DO NOT REMOVE THEM.
# new imports can be added below this line if needed.

class LEFM_QPERY_b5_2FeatureMap(BaseFeatureMap):
    """LEFM with Quadratic Post-Entanglement RY (LEFM-QPERY-b5.2) class.

    Args:
        BaseFeatureMap (_type_): base feature map class

    Example:
    """

    def __init__(self, n_qubits: int, a_coeffs: np.ndarray = None) -> None:
        """Initialize the LEFM with Quadratic Post-Entanglement RY feature map class.

        Args:
            n_qubits (int): number of qubits
            a_coeffs (np.ndarray): Coefficients for RX rotations. Default is np.ones(n_qubits).
        """
        super().__init__(PENNYLANE_PLATFORM, n_qubits)
        # hyperparameters
        self.n_qubits: int = n_qubits
        self.a_coeffs: np.ndarray = (a_coeffs if a_coeffs is not None else np.ones(n_qubits))

    def feature_map(self, x: np.ndarray) -> None:
        """Create quantum circuit of feature map.
        The input data is a sample of MNIST image data. It is decomposed into 80 features by PCA.

        Args:
            x (np.ndarray): input data shape is (80,).
        """
        n_features = x.shape[0]
        n_layers = n_features // (2 * self.n_qubits)  # Each layer uses 2*n_qubits features

        for layer in range(n_layers):
            # RX Encoding Layer
            for i in range(self.n_qubits):
                qml.RX(phi=self.a_coeffs[i] * x[layer * 2 * self.n_qubits + i], wires=i)

            # Entanglement Layer
            for i in range(self.n_qubits - 1):
                qml.CNOT(wires=[i, i + 1])
            qml.CNOT(wires=[self.n_qubits - 1, 0])

            # Quadratic RY Rotation Layer
            for i in range(self.n_qubits):
                qml.RY(phi=5.2 * x[layer * 2 * self.n_qubits + self.n_qubits + i] ** 2, wires=i)
\end{lstlisting}

\subsection{Claude by Anthropic}
The Claude series includes multiple models designed for different types of tasks. To maintain consistency with the model-task correspondence used for OpenAI’s models, we assigned specific Gemini models to each task category. In particular, ``claude-3-7-sonnet-20250219'' was used for review, idea-generation, idea-reflection, and code-generation tasks; ``claude-3-5-haiku-20241022'' was used for scoring, validation and summary tasks. The ``claude-3-7-sonnet-20250219'' model allows for adjusting the depth of reasoning by specifying the number of tokens allocated for the reasoning process. In this study, we allocated 20,000 tokens for this purpose.

We confirmed that the proposed agentic system also functions effectively when using the Claude model. Compared to other LLMs, the Claude model tended to produce more complex quantum feature maps and generate code of larger scale (Listing \ref{lst:BestGeneratedCode}, \ref{lst:GeminiGeneratedCode}, \ref{lst:ClaudeGeneratedCode}). On the other hand, some of the generated ideas included nonlinear transformations, which were explicitly prohibited in the prompts. This suggests the need for careful design of the reasoning process and prompt engineering tailored to the specific characteristics of the model.

\begin{lstlisting}[language=Python, inputencoding=utf8, style=codeFormat, caption={Generated feature map code by Claude model}, label=lst:ClaudeGeneratedCode]
import numpy as np
import pennylane as qml
from qxmt.constants import PENNYLANE_PLATFORM
from qxmt.feature_maps import BaseFeatureMap

class DualPhasePatternEncoder(BaseFeatureMap):
    """Dual-Phase Pattern Quantum Encoder feature map.

    This feature map implements different phase patterns in first versus second repetition
    while preserving complementary structures.

    Args:
        BaseFeatureMap (_type_): base feature map class

    Example:
        >>> feature_map = DualPhasePatternEncoder(n_qubits=10)
    """

    def __init__(
        self, 
        n_qubits: int, 
        scale_factor: float = 0.94 * np.pi, 
        offset: float = np.pi / 3.35,
        rep1_phase1_even: float = np.pi / 3,
        rep1_phase1_odd: float = np.pi / 2,
        rep1_phase2_even: float = np.pi / 4,
        rep1_phase2_odd: float = np.pi / 2,
        rep1_phase3_all: float = np.pi / 4,
        rep2_phase1_even: float = np.pi / 2,
        rep2_phase1_odd: float = np.pi / 3,
        rep2_phase2_even: float = np.pi / 2,
        rep2_phase2_odd: float = np.pi / 4,
        rep2_phase3_all: float = np.pi / 3,
        h_mod_phase1: float = np.pi / 6,
        h_mod_phase2: float = np.pi / 2,
        h_mod_phase3: float = 5 * np.pi / 6,
        reps: int = 2
    ) -> None:
        """Initialize the Dual-Phase Pattern Quantum Encoder feature map.

        Args:
            n_qubits (int): number of qubits
            scale_factor (float, optional): Scaling factor for feature angles. Defaults to 0.94*np.pi.
            offset (float, optional): Offset for feature angles. Defaults to np.pi/3.35.
            rep1_phase1_even (float, optional): Phase for even qubits after layer 1 in rep 1. Defaults to π/3.
            rep1_phase1_odd (float, optional): Phase for odd qubits after layer 1 in rep 1. Defaults to π/2.
            rep1_phase2_even (float, optional): Phase for even qubits after layer 2 in rep 1. Defaults to π/4.
            rep1_phase2_odd (float, optional): Phase for odd qubits after layer 2 in rep 1. Defaults to π/2.
            rep1_phase3_all (float, optional): Phase for all qubits after layer 3 in rep 1. Defaults to π/4.
            rep2_phase1_even (float, optional): Phase for even qubits after layer 1 in rep 2. Defaults to π/2.
            rep2_phase1_odd (float, optional): Phase for odd qubits after layer 1 in rep 2. Defaults to π/3.
            rep2_phase2_even (float, optional): Phase for even qubits after layer 2 in rep 2. Defaults to π/2.
            rep2_phase2_odd (float, optional): Phase for odd qubits after layer 2 in rep 2. Defaults to π/4.
            rep2_phase3_all (float, optional): Phase for all qubits after layer 3 in rep 2. Defaults to π/3.
            h_mod_phase1 (float, optional): Phase for mod 4 = 1 qubits before Hadamard. Defaults to π/6.
            h_mod_phase2 (float, optional): Phase for mod 4 = 2 qubits before Hadamard. Defaults to π/2.
            h_mod_phase3 (float, optional): Phase for mod 4 = 3 qubits before Hadamard. Defaults to 5π/6.
            reps (int, optional): Number of repetitions. Defaults to 2.
        """
        super().__init__(PENNYLANE_PLATFORM, n_qubits)
        # hyperparameters
        self.n_qubits: int = n_qubits
        self.scale_factor: float = scale_factor
        self.offset: float = offset
        
        # Phase pattern for first repetition
        self.rep1_phase1_even: float = rep1_phase1_even
        self.rep1_phase1_odd: float = rep1_phase1_odd
        self.rep1_phase2_even: float = rep1_phase2_even
        self.rep1_phase2_odd: float = rep1_phase2_odd
        self.rep1_phase3_all: float = rep1_phase3_all
        
        # Phase pattern for second repetition
        self.rep2_phase1_even: float = rep2_phase1_even
        self.rep2_phase1_odd: float = rep2_phase1_odd
        self.rep2_phase2_even: float = rep2_phase2_even
        self.rep2_phase2_odd: float = rep2_phase2_odd
        self.rep2_phase3_all: float = rep2_phase3_all
        
        # Hadamard pattern phases
        self.h_mod_phase1: float = h_mod_phase1
        self.h_mod_phase2: float = h_mod_phase2
        self.h_mod_phase3: float = h_mod_phase3
        self.reps: int = reps
        
        # Define triplets for controlled-Z gates
        self.cz_triplets = [
            (0, 3, 6), (1, 4, 7), (2, 5, 8), (3, 6, 9), (4, 7, 0),
            (5, 8, 1), (6, 9, 2), (7, 0, 3), (8, 1, 4), (9, 2, 5)
        ]

    def _encode_features_first_rep(self, x: np.ndarray) -> None:
        """Apply feature encoding for the first repetition.
        
        First repetition (30 features):
        * Features 1-6 → Rx rotations on qubits 1-6
        * Features 7-24 → Ry rotations on qubits 7-10 and 1-14 (exactly 18 Ry gates)
        * Features 25-30 → Rz rotations on qubits 5-10
        
        Args:
            x (np.ndarray): Input data of shape (80,)
        """
        # Features 1-6 → Rx rotations on qubits 1-6 (0-5 in 0-indexed)
        for i in range(min(6, len(x))):
            angle = self.scale_factor * x[i] + self.offset
            qml.RX(phi=angle, wires=i % self.n_qubits)
        
        # Features 7-24 → Ry rotations on qubits 7-10 and 1-14 (exactly 18 Ry gates)
        for i in range(6, min(24, len(x))):
            angle = self.scale_factor * x[i] + self.offset
            # Map to qubits 7-10 (6-9 in 0-indexed) and then 1-14 (0-13 in 0-indexed)
            if i < 10:  # For features 7-10 map to qubits 7-10 (6-9 in 0-indexed)
                wire_idx = i
            else:  # For features 11-24 map to qubits 1-14 (0-13 in 0-indexed)
                wire_idx = (i - 10) % self.n_qubits
            qml.RY(phi=angle, wires=wire_idx % self.n_qubits)
        
        # Features 25-30 → Rz rotations on qubits 5-10 (4-9 in 0-indexed)
        for i in range(24, min(30, len(x))):
            angle = self.scale_factor * x[i] + self.offset
            wire_idx = (i - 24 + 4) % self.n_qubits  # Map to qubits 5-10 (4-9 in 0-indexed)
            qml.RZ(phi=angle, wires=wire_idx)

    def _encode_features_second_rep(self, x: np.ndarray) -> None:
        """Apply feature encoding for the second repetition.
        
        Second repetition (30 features):
        * Features 31-36 → Rx rotations on qubits 5-10
        * Features 37-54 → Ry rotations on qubits 1-18 (exactly 18 Ry gates)
        * Features 55-60 → Rz rotations on qubits 1-6
        
        Args:
            x (np.ndarray): Input data of shape (80,)
        """
        # Features 31-36 → Rx rotations on qubits 5-10 (4-9 in 0-indexed)
        for i in range(30, min(36, len(x))):
            angle = self.scale_factor * x[i] + self.offset
            wire_idx = (i - 30 + 4) % self.n_qubits  # Map to qubits 5-10 (4-9 in 0-indexed)
            qml.RX(phi=angle, wires=wire_idx)
        
        # Features 37-54 → Ry rotations on qubits 1-18 (0-17 in 0-indexed)
        for i in range(36, min(54, len(x))):
            angle = self.scale_factor * x[i] + self.offset
            wire_idx = (i - 36) % self.n_qubits  # Map to qubits 1-18 (0-17 in 0-indexed)
            qml.RY(phi=angle, wires=wire_idx)
        
        # Features 55-60 → Rz rotations on qubits 1-6 (0-5 in 0-indexed)
        for i in range(54, min(60, len(x))):
            angle = self.scale_factor * x[i] + self.offset
            wire_idx = (i - 54) % self.n_qubits  # Map to qubits 1-6 (0-5 in 0-indexed)
            qml.RZ(phi=angle, wires=wire_idx)

    def _encode_final_layer(self, x: np.ndarray) -> None:
        """Apply feature encoding for the final layer.
        
        Final encoding layer (20 features):
        * Features 61-64 → Rx rotations on qubits 7-10
        * Features 65-80 → Ry rotations on qubits 1-16 (exactly 16 Ry gates)
        
        Args:
            x (np.ndarray): Input data of shape (80,)
        """
        # Features 61-64 → Rx rotations on qubits 7-10 (6-9 in 0-indexed)
        for i in range(60, min(64, len(x))):
            angle = self.scale_factor * x[i] + self.offset
            wire_idx = (i - 60 + 6) % self.n_qubits  # Map to qubits 7-10 (6-9 in 0-indexed)
            qml.RX(phi=angle, wires=wire_idx)
        
        # Features 65-80 → Ry rotations on qubits 1-16 (0-15 in 0-indexed)
        for i in range(64, min(80, len(x))):
            angle = self.scale_factor * x[i] + self.offset
            wire_idx = (i - 64) % self.n_qubits  # Map to qubits 1-16 (0-15 in 0-indexed)
            qml.RY(phi=angle, wires=wire_idx)

    def _apply_local_entanglement(self) -> None:
        """Apply CNOT gates between adjacent qubits (Layer 1)."""
        for i in range(self.n_qubits):
            qml.CNOT(wires=[i, (i + 1) % self.n_qubits])
    
    def _apply_medium_entanglement(self) -> None:
        """Apply CNOT gates between qubits separated by distance 2 (Layer 2)."""
        for i in range(self.n_qubits):
            qml.CNOT(wires=[i, (i + 2) % self.n_qubits])
    
    def _apply_global_entanglement(self) -> None:
        """Apply CNOT gates between qubits separated by distance n/3 (Layer 3)."""
        distance = max(1, self.n_qubits // 3)  # Ensure distance is at least 1
        for i in range(self.n_qubits):
            qml.CNOT(wires=[i, (i + distance) % self.n_qubits])
    
    def _apply_phase1_rep1(self) -> None:
        """Apply Phase pattern after Layer 1 in repetition 1:
        Rz(π/3) to even-indexed qubits and Rz(π/2) to odd-indexed qubits."""
        for i in range(self.n_qubits):
            if i % 2 == 0:  # even
                qml.RZ(phi=self.rep1_phase1_even, wires=i)
            else:  # odd
                qml.RZ(phi=self.rep1_phase1_odd, wires=i)
    
    def _apply_phase2_rep1(self) -> None:
        """Apply Phase pattern after Layer 2 in repetition 1:
        Rz(π/4) to even-indexed qubits and Rz(π/2) to odd-indexed qubits."""
        for i in range(self.n_qubits):
            if i % 2 == 0:  # even
                qml.RZ(phi=self.rep1_phase2_even, wires=i)
            else:  # odd
                qml.RZ(phi=self.rep1_phase2_odd, wires=i)
    
    def _apply_phase3_rep1(self) -> None:
        """Apply Phase pattern after Layer 3 in repetition 1:
        Rz(π/4) to all qubits."""
        for i in range(self.n_qubits):
            qml.RZ(phi=self.rep1_phase3_all, wires=i)
    
    def _apply_phase1_rep2(self) -> None:
        """Apply Phase pattern after Layer 1 in repetition 2:
        Rz(π/2) to even-indexed qubits and Rz(π/3) to odd-indexed qubits."""
        for i in range(self.n_qubits):
            if i % 2 == 0:  # even
                qml.RZ(phi=self.rep2_phase1_even, wires=i)
            else:  # odd
                qml.RZ(phi=self.rep2_phase1_odd, wires=i)
    
    def _apply_phase2_rep2(self) -> None:
        """Apply Phase pattern after Layer 2 in repetition 2:
        Rz(π/2) to even-indexed qubits and Rz(π/4) to odd-indexed qubits."""
        for i in range(self.n_qubits):
            if i % 2 == 0:  # even
                qml.RZ(phi=self.rep2_phase2_even, wires=i)
            else:  # odd
                qml.RZ(phi=self.rep2_phase2_odd, wires=i)
    
    def _apply_phase3_rep2(self) -> None:
        """Apply Phase pattern after Layer 3 in repetition 2:
        Rz(π/3) to all qubits."""
        for i in range(self.n_qubits):
            qml.RZ(phi=self.rep2_phase3_all, wires=i)
    
    def _apply_cz_triplets(self) -> None:
        """Apply controlled-Z gates to strategic triplets."""
        for a, b, c in self.cz_triplets:
            if a < self.n_qubits and b < self.n_qubits and c < self.n_qubits:
                qml.CZ(wires=[a, b])
                qml.CZ(wires=[b, c])
                qml.CZ(wires=[c, a])
    
    def _apply_fourier_hadamard(self) -> None:
        """Apply Fourier-Inspired Hadamard Pattern:
        - Qubit index mod 4 = 0: Apply H gate
        - Qubit index mod 4 = 1: Apply Rz(π/6) followed by H gate
        - Qubit index mod 4 = 2: Apply Rz(π/2) followed by H gate
        - Qubit index mod 4 = 3: Apply Rz(5π/6) followed by H gate
        """
        for i in range(self.n_qubits):
            mod4 = i % 4
            if mod4 == 0:
                qml.Hadamard(wires=i)
            elif mod4 == 1:
                qml.RZ(phi=self.h_mod_phase1, wires=i)
                qml.Hadamard(wires=i)
            elif mod4 == 2:
                qml.RZ(phi=self.h_mod_phase2, wires=i)
                qml.Hadamard(wires=i)
            elif mod4 == 3:
                qml.RZ(phi=self.h_mod_phase3, wires=i)
                qml.Hadamard(wires=i)
    
    def feature_map(self, x: np.ndarray) -> None:
        """Create quantum circuit of feature map.
        The input data is a sample of MNIST image data. It is decomposed into 80 features by PCA.
        
        Args:
            x (np.ndarray): input data shape is (80,).
        """
        # First repetition with its specific phase pattern
        self._encode_features_first_rep(x)
        
        # Apply entanglement layers with phase pattern for first repetition
        self._apply_local_entanglement()
        self._apply_phase1_rep1()
        
        self._apply_medium_entanglement()
        self._apply_phase2_rep1()
        
        self._apply_global_entanglement()
        self._apply_phase3_rep1()
        
        # Apply controlled-Z triplet pattern
        self._apply_cz_triplets()
        
        # Second repetition with its specific phase pattern
        self._encode_features_second_rep(x)
        
        # Apply entanglement layers with phase pattern for second repetition
        self._apply_local_entanglement()
        self._apply_phase1_rep2()
        
        self._apply_medium_entanglement()
        self._apply_phase2_rep2()
        
        self._apply_global_entanglement()
        self._apply_phase3_rep2()
        
        # Apply controlled-Z triplet pattern
        self._apply_cz_triplets()
        
        # Apply final encoding layer
        self._encode_final_layer(x)
        
        # Apply Fourier-Inspired Hadamard Pattern
        self._apply_fourier_hadamard()
\end{lstlisting}

\section{Prompts}\label{sec:prompts}
In the developed agentic system, a total of eight types of prompts are used to instruct the LLM. These prompts are explained separately by component. At runtime, they may receive multiple placeholders (represented in in red color text), the details of which are provided in the Table~\ref{tab:promptPlaceholder}. In the prompt, the terms ``Trial'' and ``Round'' both refer to repetitions, but they denote different levels of iteration: a ``Trial'' represents a complete cycle of the agentic system—including Generation, Validation, Evaluation, and Review—while a ``Round'' refers to repeated processes within specific components of a single Trial, such as Scouring or Reflection.

\subsection{Generation}
\label{sec:methodsGenerationPrompts}

In the ``Generation'' component, five types of prompts are used: idea generation, idea scoring, idea reflection, summary generation, and code generation.  
The corresponding prompt files are shown in Listings~\ref{lst:IdeaGenerationDeveloper}--\ref{lst:BaseFeatureMapCode}.

\begin{itemize}
    \item Idea generation: Listings~\ref{lst:IdeaGenerationDeveloper}--\ref{lst:IdeaGenerationUserSubsequent}
    \item Idea scoring: Listings~\ref{lst:IdeaScoringDeveloper}--\ref{lst:IdeaScoringUserFinal}
    \item Idea reflection: Listings~\ref{lst:IdeaReflectionDeveloper}--\ref{lst:IdeaReflectionUser}
    \item Summary generation: Listings~\ref{lst:SummarizationDeveloper}--\ref{lst:SummarizationUser}
    \item Code generation: Listings~\ref{lst:CodeGenerationDeveloper}--\ref{lst:BaseFeatureMapCode}
\end{itemize}

\begin{lstlisting}[style=promptFormat, caption={The developer prompt for the idea generation}, label=lst:IdeaGenerationDeveloper]
"""
You are a quantum computing expert specializing in designing Quantum Feature Maps for classification tasks using a Quantum Support Vector Machine (QSVM).

Your task is to create a quantum feature map that will serve as the kernel function in a QSVM classifier applied to MNIST data. The ultimate goal is to design a feature map that enables the classifier to achieve high accuracy in classification.

This task will follow an iterative improvement process, where the feature map design is refined based on review comments provided after each iteration. Use the feedback to enhance the design while maintaining alignment with the defined objectives and constraints.

# Task Definition
Develop multiple ideas for quantum feature maps that satisfy the following criteria. The feature maps will be used to compute the quantum kernel K(x, x') = |⟨Φ(x)|Φ(x')⟩|^2 for a QSVM. Ensure the designs are tailored to this kernel computation.
1. **Design Considerations**:
    - Define combinations of quantum gates.
    - Define an entanglement pattern for the features.
    - Specify the method for embedding input data and quantum states as rotation angles of quantum gates.
    - Ensure the 80-dimensional input data is utilized effectively, minimizing any loss of information.
        - Avoid excessive feature compression that may lead to information loss (e.g., simple feature averaging, summing, etc.).
2. **Restrictions on Encoding and Embedding**:
    - Only **linear functions** are allowed for encoding and embedding.
    - All parameters in the encoding and embedding must be **non-trainable**.

## Key Context
- The input data, originally represented as 784-dimensional image data, has been compressed to 80 dimensions using PCA and each value normalized to the range [0.0, 1.0].
- Propose a quantum feature map that is independent of the number of qubits in the quantum device.

## Iterative Design
- You will refine this feature map over {max_trial_num} total trials.
- Each trial, you'll receive evaluation feedback to help evolve the design.
- In subsequent trials, the primary goal is to improve classification accuracy based on the feedback provided, while maintaining or improving the feature map's fidelity and computational feasibility.

## Output Format
results: [idea_1, idea_2, ..., idea_n]
Each idea should be structured as follows:
    - explanation: A detailed explanation of the proposed feature map. Include design rationale, expected outcomes, and quantum gates used.
    - formula: A concise TeX-formatted mathematical representation of your idea.
    - summary: A 100-300 word summary highlighting the core innovation.
    - feature_map_name: A descriptive name for your feature map.
    - key_sentences: Up to 5 key sentences, each 50-100 words, that describe the essential aspects of your design for subsequent vector-based searching.

## Important Notes
- Clarity is paramount. Reiterate points if necessary to ensure understanding. There is no length restriction on the explanation-ensure all relevant details are provided.
- Evaluation is performed using an ideal quantum simulator without noise, so hardware noise does not need to be considered.
"""
\end{lstlisting}
\begin{lstlisting}[style=promptFormat, caption={The user prompt for the idea generation (first trial)}, label=lst:IdeaGenerationUserFirst]
"""
Trial {current_trial}/{max_trial_num}.

This is the **first trial** of the quantum feature map design task.  
For this initial trial:  
- Focus on designing high-accuracy quantum feature maps while exploring diverse approaches to feature map design.
- Ensure the designs align with the following constraints:
    1. The idea itself must be a feature map that is independent of the number of qubits in the quantum device, but evaluation will be conducted using an {device_n_qubit}-qubit simulator.
    2. The input data consists of **80-dimensional PCA-reduced MNIST data**, with each value normalized to the range [0.0, 1.0].
    3. The encoding method is restricted to **non-trainable parameters** and **linear functions**.
    4. Ensure effective utilization of the 80-dimensional input data while minimizing information loss.
        - Avoid excessive feature compression that may lead to information loss (e.g., simple feature averaging, summing, etc.).

### Key Objective
Create **{idea_num} high-accuracy quantum feature map ideas** that explore diverse directions and can serve as strong foundations for refinement in future trials.
"""
\end{lstlisting}
\begin{lstlisting}[style=promptFormat, caption={The user prompt for the idea generation (subsequent trials)}, label=lst:IdeaGenerationUserSubsequent]
"""
Trial {current_trial}/{max_trial_num}.

### Feedback from the Previous Trial (Trial {previous_trial})
{review_comment}

### Task for This Trial
Based on the feedback provided above, refine your quantum feature map ideas and generate a total of **{idea_num} improved ideas**. These ideas should aim to enhance the classification accuracy.

### Key Guidelines
- The feedback includes the following:  
    - **Keep Points:** Aspects of the previous design that should be retained.  
    - **Suggestions:** Areas for improvement or new directions to explore.  

- Variety in Approaches:  
    - You are not required to address all feedback points in a single idea.
    - Ensure that the multiple ideas generated based on the review incorporate different directions of improvement, maintaining diversity.

### Primary Objective
The primary goal in this trial is to **boost classification accuracy** through iterative refinements while ensuring the designs align with the constraints.
"""
\end{lstlisting}

\begin{lstlisting}[style=promptFormat, caption={The developer prompt for the idea scoring}, label=lst:IdeaScoringDeveloper]
"""
As a reviewer for a scientific journal, you are tasked with evaluating new scientific ideas from multiple perspectives while adhering to specific evaluation criteria. Your evaluation should be thorough, objective, and based on the guidelines provided below.

# Evaluation Criteria
Each criterion is scored on a scale of 0.0 to 10.0, in increments of 0.1, where 0.0 represents the lowest score and 10.0 represents the highest score:
- Originality: Assess how the idea differs from existing research. Does it make a novel contribution?
- Feasibility: Evaluate the practicality of implementing the idea.
- Versatility: Consider how broadly the idea can be applied.

# Steps for Evaluation
1. Understand the Idea: 
    - Carefully read and comprehend the proposed idea.
    - Organize the information needed to make a well-informed evaluation.
2. Assess Information Sufficiency: 
    - First, confirm the round number provided by the user. If it is the final round, skip Step 2 and proceed to Step 3.
    - Otherwise, assess whether the "# Related Work" section provides sufficient information to evaluate the idea.
    - If the information is insufficient for scoring, <is_lack_information> tag set to True, and list up to 5 necessary information key sentences as a comma-separated list within <additional_key_sentences> tags. The search will be conducted by embedded vector for academic paper; therefore, ensure the <additional_key_sentences> are specific and relevant. Each sentence length should be between 50 to 100 words.
    - In this case, terminate the scoring process and set all scores to 0.0.
3. Provide Reasoning:
    - If the information is sufficient, proceed to evaluate the idea based on the specified criteria.
    - Enclose the rationale behind the evaluation results of each indicator in <reason> tags and explain it in text.
4. Assign Scores:
    - Based on the evaluation results and their rationale, assign a score to each indicator.
5. Terminating the Evaluation:
    - Once all scores and reasoning have been assigned, the evaluation is complete. <is_lack_information> tag set to False.

# Baseline
{few_shot_examples}
"""
\end{lstlisting}
\lstinputlisting[style=promptFormat, caption={The few shots example for idea scoring}, label=lst:IdeaScoringFewShots]{prompts/idea_scoring_few_shots.py}
\begin{lstlisting}[style=promptFormat, caption={The user prompt for the idea scoring (first round)}, label=lst:IdeaScoringUserFirst]
"""
Round {current_round}/{max_scoring_round}.

You are tasked with evaluating the following "# Proposed Idea" using the specified criteria and providing scores for each criterion. The idea represents a newly proposed quantum feature map for the quantum kernel method. Your evaluation and scoring should consider multiple perspectives and adhere to the strictest possible standards. Finally, after all the scores have been assigned, summarize the rationale for each score.

# Proposed Idea
{idea}

# Related Work
{related_work}
"""
\end{lstlisting}
\begin{lstlisting}[style=promptFormat, caption={The user prompt for the idea scoring (intermediate rounds)}, label=lst:IdeaScoringUserIntermediate]
"""
Round {current_round}/{max_scoring_round}.

Additional Related Work has been provided for further context. Please evaluate the proposed idea using the specified criteria and provide scores for each criterion.

# Related Work
{related_work}
"""
\end{lstlisting}
\begin{lstlisting}[style=promptFormat, caption={The user prompt for the idea scoring (final rounds)}, label=lst:IdeaScoringUserFinal]
"""
Round {current_round}/{max_scoring_round} (Final Round).

Additional Related Work has been provided for further context.

This marks the final opportunity to request additional information. Based on the provided details, conduct a comprehensive evaluation and ensure scores are assigned to each criterion without exception.
This round DOES NOT return "is_lack_information"=True. You must provide scores for each criterion.

# Related Work
{related_work}
"""
\end{lstlisting}

\begin{lstlisting}[style=promptFormat, caption={The developer prompt for the idea reflection}, label=lst:IdeaReflectionDeveloper]
"""
You are a professor with extensive expertise in **quantum computing** and **machine learning**, particularly in scientific research.

Your task is to **evaluate quantum machine learning ideas** from multiple perspectives, incorporating insights from recent academic papers and best practices. Focus on refining each idea to improve its accuracy and effectiveness based on the latest advancements, while preserving the idea's core structure.

### Notes
- **Incorporate recent advancements:** Utilize relevant developments in quantum technology and machine learning where applicable to enhance the evaluation.
- **Provide balanced evaluations:** While not all perspectives will apply equally to every idea, strive to deliver a thorough and well-rounded assessment.
- **Design considerations:** Restrict encoding to **non-trainable parameters** and **linear functions**.
- **Focus on research-supported improvements:** Suggest refinements supported by related research, without making fundamental changes to the core concept of the idea.
- **Retain original content when no changes are required:** If no modifications are needed, keep the content of each tag unchanged and include it as-is in the output.
"""
\end{lstlisting}
\begin{lstlisting}[style=promptFormat, caption={The user prompt for the idea reflection}, label=lst:IdeaReflectionUser]
"""
Round {current_round}/{max_reflection_round}.

Carefully review the idea provided in the "# Previous Idea" section, along with its score from the "# First Round Score" section. When conducting your evaluation, take into account the relevant academic papers and insights listed in the "# Related Work" section. 

After your analysis and evaluation:
- Refine and improve the idea for high-accuracy where appropriate, ensuring that the **core concept remains unchanged**.
- If no modifications are required, retain the following tags as-is: `feature_map_name`, `summary`, `explanation`, `formula`, and `key_sentences`.
- In cases where no changes are necessary, set the `is_completed` tag to **True** in your output.

# Previous Idea
{previous_idea}

# First Round Score:
{previous_score}

# Related Work
{related_work}
"""
\end{lstlisting}

\begin{lstlisting}[style=promptFormat, caption={The developer prompt for the summary generation}, label=lst:SummarizationDeveloper]
"""
You are an academic journal editor. Your task is to thoroughly understand the full content of the paper provided by the user and summarize it. When summarizing, rely solely on the information from the provided paper and avoid referencing external sources. Ensure that the summary accurately reflects the authors' arguments and claims. Write a detailed summary of approximately {max_summary_words} words.
"""
\end{lstlisting}
\begin{lstlisting}[style=promptFormat, caption={The user prompt for the summary generation}, label=lst:SummarizationUser]
"""
Follow these steps to create a summary of the paper:
1. The full text of the paper is provided in the section titled "## Full content of paper." Carefully read and understand its content.
2. Create a detailed summary with approximately {max_summary_words} words, focusing on the following aspects: 
    - Key findings
    - Methodology
    - Results
    - Future works or potential areas for improvement

## Full content of paper
{raw_content}
"""
\end{lstlisting}

\begin{lstlisting}[style=promptFormat, caption={The developer prompt for the code generation}, label=lst:CodeGenerationDeveloper]
"""
You are a quantum computing expert specializing in designing Quantum Feature Maps for classification tasks using a Quantum Support Vector Machine (QSVM).

Your task is to create a quantum feature map that will serve as the kernel function in a QSVM classifier applied to MNIST data. The ultimate goal is to design a feature map that enables the classifier to achieve high accuracy in classification.

This task will follow an iterative improvement process, where the feature map design is refined based on review comments provided after each iteration. Use the feedback to enhance the design while maintaining alignment with the defined objectives and constraints.

# Task Definition
Develop multiple ideas for quantum feature maps that satisfy the following criteria. The feature maps will be used to compute the quantum kernel K(x, x') = |⟨Φ(x)|Φ(x')⟩|^2 for a QSVM. Ensure the designs are tailored to this kernel computation.
1. **Design Considerations**:
    - Define combinations of quantum gates.
    - Define an entanglement pattern for the features.
    - Specify the method for embedding input data and quantum states as rotation angles of quantum gates.
    - Ensure the 80-dimensional input data is utilized effectively, minimizing any loss of information.
        - Avoid excessive feature compression that may lead to information loss (e.g., simple feature averaging, summing, etc.).
2. **Restrictions on Encoding and Embedding**:
    - Only **linear functions** are allowed for encoding and embedding.
    - All parameters in the encoding and embedding must be **non-trainable**.

## Key Context
- The input data, originally represented as 784-dimensional image data, has been compressed to 80 dimensions using PCA and each value normalized to the range [0.0, 1.0].
- Propose a quantum feature map that is independent of the number of qubits in the quantum device.

## Iterative Design
- You will refine this feature map over {max_trial_num} total trials.
- Each trial, you'll receive evaluation feedback to help evolve the design.
- In subsequent trials, the primary goal is to improve classification accuracy based on the feedback provided, while maintaining or improving the feature map's fidelity and computational feasibility.

## Output Format
results: [idea_1, idea_2, ..., idea_n]
Each idea should be structured as follows:
    - explanation: A detailed explanation of the proposed feature map. Include design rationale, expected outcomes, and quantum gates used.
    - formula: A concise TeX-formatted mathematical representation of your idea.
    - summary: A 100-300 word summary highlighting the core innovation.
    - name: A descriptive name for your feature map.
    - key_sentences: Up to 5 key sentences, each 50-100 words, that describe the essential aspects of your design for subsequent vector-based searching.

## Important Notes
- Clarity is paramount. Reiterate points if necessary to ensure understanding. There is no length restriction on the explanation-ensure all relevant details are provided.
- Evaluation is performed using an ideal quantum simulator without noise, so hardware noise does not need to be considered.
"""
\end{lstlisting}
\begin{lstlisting}[style=promptFormat, caption={The user prompt for the code generation}, label=lst:CodeGenerationUser]
"""
Please implement a quantum feature map based on the designs described in the "# Idea" section by extending the `BaseFeatureMap` class using PennyLane code. The idea is based on the assumption that it does not depend on the number of qubits. However, the generated code will be executed on a 10-qubit simulator.

# Idea
{idea}
"""
\end{lstlisting}
\begin{lstlisting}[style=codeFormat, caption={The base code for implementing generated feature map idea}, label=lst:BaseFeatureMapCode]
import numpy as np
import pennylane as qml
from qxmt.constants import PENNYLANE_PLATFORM
from qxmt.feature_maps import BaseFeatureMap

# above are the default imports. DO NOT REMOVE THEM.
# new imports can be added below this line if needed.

class SeedFeatureMap(BaseFeatureMap):
    """Seed feature map class.

    Args:
        BaseFeatureMap (_type_): base feature map class

    Example:
    """

    def __init__(self, n_qubits: int) -> None:
        """ "Initialize the Seed feature map class.

        Args:
            n_qubits (int): number of qubits
        """
        super().__init__(PENNYLANE_PLATFORM, n_qubits)
        # hyperparameters
        self.n_qubits: int = n_qubits

    def feature_map(self, x: np.ndarray) -> None:
        """Create quantum circuit of feature map.
        The input data is a sample of MNIST image data. It is decomposed into 80 features by PCA.

        Args:
            x (np.ndarray): input data shape is (80,).
        """
        # define your quantum feature map here
        pass
\end{lstlisting}

\subsection{Validation}
\label{sec:methodsValidationPrompt}

In the ``Validation'' component, the prompts were created to check whether the generated code appropriately uses the PennyLane library methods, and to assist in correcting any errors.  
The relevant listings are shown in Listings~\ref{lst:CodeValidationDeveloper}--\ref{lst:ErrorCorrectingUser}.

\begin{itemize}
    \item Code validation: Listings~\ref{lst:CodeValidationDeveloper}--\ref{lst:CodeValidationUser}
    \item Error correcting: Listings~\ref{lst:ErrorCorrectingDeveloper}--\ref{lst:ErrorCorrectingUser}
\end{itemize}

\begin{lstlisting}[style=promptFormat, caption={The developer prompt for the code validation}, label=lst:CodeValidationDeveloper]
"""
You are an expert quantum software engineer especially skilled in the PennyLane library.

Your Task is to extracted argments from the user provided PennyLane function and PennyLane documentation. The Details are as follows:
# Task
For each function or class in the provided input, follow these steps:

1. **Extract the Class Name**:
    - Identify the function or class name from the user-provided code in the "# User Provided PennyLane Class and Method" section.
    - The class name starts with `qml.` and ends before the first `(`.
    - Do not include the parentheses or any characters after them.
2. **Extract User-Defined Argument Names**:
    - Identify argument names in the user code based on the patterns described in the "# Expected User Arguments Pattern" section.
    - For **Pattern 1**, arguments are values only (no `arg_name` or `=`). This pattern is excluded from validation and should be ignored in this step.
    - For **Pattern 2** and **Pattern 3**, extract the `arg_name` portion only. If multiple arguments are present, separate them by commas and exclude type hints, values, and assignment operators (`=`).
3. **Extract Reference Argument Names from Documentation**:
    - Refer to the corresponding "Class Name" section in the "# PennyLane Documentation" part.
    - The arguments are listed as `"name" ("type"): "description"`. Only extract the `name` portion.
    - If multiple arguments exist, separate them by commas.

# Expected User Arguments Pattern
- Pattern 1: 
    - qml.Hoge(`value`) => Expected: ignore
    - qml.Hoge(`value_1`, `value_2`) => Expected: ignore
- Pattern 2: 
    - qml.Hoge(`arg_name = value`) => Expected: arg_name
    - qml.Hoge(`arg_name=value`) => Expected: arg_name
    - qml.Hoge(`arg_name_1=value_1`, `arg_name_2=value_2`) => Expected: arg_name_1, arg_name_2
    - qml.Hoge(`value_1`, `arg_name_2=value_2`) => Expected: arg_name_2
- Pattern 3:
    - qml.Hoge(`arg_name: arg_type = value`) => Expected: arg_name
    - qml.Hoge(`arg_name:arg_type=value`) => Expected: arg_name
    - qml.Hoge(`arg_name_1:arg_type=value_1`, `arg_name_2:arg_type=value_2`) => Expected: arg_name_1, arg_name_2
    - qml.Hoge(`value_1`, `arg_name_2:arg_type=value_2`) => Expected: arg_name_2

# Output JSON Format
``` json
[
    {{
        "class_name": "Name of the user provided PennyLane class (step1)",
        "user_args_name": "Argment name list that extracted user code (step2)",
        "docs_args_name": "Argument name list that extracted pennylane documentation (step3)"
    }},
    // Repeat for each case
]
```
"""
\end{lstlisting}
\begin{lstlisting}[style=promptFormat, caption={The user prompt for the code validation}, label=lst:CodeValidationUser]
"""
Pennylane method and class list is provided in "# User Provided PennyLane Class and Method" section. PennyLane documentation is provided in "# PennyLane Documentation" section. Plese extract the argments from the user provided PennyLane method and PennyLane documentation.

Make sure the output format is defined in "# Output JSON Format" section.

# User Provided PennyLane Class and Method
{methods}

# PennyLane Documentation
{references}
"""
\end{lstlisting}

\begin{lstlisting}[style=promptFormat, caption={The developer prompt for the error correcting}, label=lst:ErrorCorrectingDeveloper]
"""
You are an expert qunatum computing software engineer skilled in designing quantum feature maps using the PennyLane library in Python.

## Task Definition
You are tasked with fixing the errors and warnings in the quantum feature map code provided by the user.
Your task is only fix errors abd warnings. Do not change the logic or structure of the code.
"""
\end{lstlisting}
\begin{lstlisting}[style=promptFormat, caption={The user prompt for the error correcting}, label=lst:ErrorCorrectingUser]
"""
Please correct the following errors in the quantum feature map code.

## Code:
{code}

## Errors
{error_messages_string}

## Warnings
{warning_messages_string}

# Output Format
Make sure the code follows the same structure as the base code provided and is formatted in the following JSON format:
- class_name: Name of the generated feature map class
- params: dictionary of parameters for the feature map class
- code: generated feature map code
"""
\end{lstlisting}

\subsection{Review}
\label{sec:methodsReviewPrompt}

In the ``Review'' component, prompts are used to determine the next direction based on the ideas and the execution results.  
The related listings are shown in Listings~\ref{lst:ReviewDeveloper} and \ref{lst:ReviewUser}.

\begin{lstlisting}[style=promptFormat, caption={The developer prompt for the review}, label=lst:ReviewDeveloper]
"""
You are an expert in quantum physics and quantum machine learning, specializing in quantum feature map design.

# Task
Your task is to review past ideas on quantum feature maps and their evaluation results, and propose improvements to enhance accuracy through continuous refinement. Restrict your review to the quantum feature map design itself; do not propose changes to the overall model, evaluation metrics, or other workflows.

# Quantum Feature Map Definition
Quantum feature maps (Φ(x)) will be used to compute the quantum kernel K(x, x') = |⟨Φ(x)|Φ(x')⟩|^2 for a QSVM (Quantum Support Vector Machine).
1. **Design Considerations**:
    - Define combinations of quantum gates.
    - Define an entanglement pattern for the features.
    - Specify the method for embedding input data and quantum states as rotation angles of quantum gates.
    - Ensure the 80-dimensional input data is utilized effectively, minimizing any loss of information.
        - Avoid excessive feature compression that may lead to information loss (e.g., simple feature averaging, summing, etc.).
2. **Restrictions on Encoding and Embedding**:
    - Only **linear functions** are allowed for encoding and embedding.
    - All parameters in the encoding and embedding must be **non-trainable**.

# Output Format
1. Keep Points:
    - Identify factors that contributed to improved accuracy.
    - Analyze the most accurate idea in detail.
    - Review multiple ideas to identify common elements that contributed to improving accuracy.
2. Suggestions:
    - Limit the number of suggestions to {max_suggestion_num} or fewer.
    - Ensure each suggestion includes only a single proposal.
    - Prioritize the most impactful suggestions.
    - If no suggestions for improvement are identified, return suggestions: ["COMPLETED"].
3. Output Schema: {{
    "keep_points": ["point 1", "point 2", ..., "point n"],
    "suggestions": ["suggestion 1", "suggestion 2", ..., "suggestion n"]
    }}.

# Notes
- Input Data: The input data, originally represented as 784-dimensional image data, has been compressed to 80 dimensions using PCA and each value normalized to the range [0.0, 1.0]
- Simulation: Use an ideal quantum simulator without noise for evaluation
- Evaluation Metric: Classification accuracy is the primary metric. The goal is to achieve the highest possible accuracy.
"""
\end{lstlisting}
\begin{lstlisting}[style=promptFormat, caption={The user prompt for the review}, label=lst:ReviewUser]
"""
The previous idea and experimental results are provided below in the "# Previous Trial Idea and Results" section. These reflect iterative adjustments based on past trial review comments. 

Review the trial results to design a quantum feature map for a more accurate QSVM. Identify the factors that contributed to accuracy improvement and areas for further enhancement. Finally, check whether the review results comply with the design rules for the quantum feature map.

# Previous Trial Idea and Results (Trial Number: {last_trial_num})
{last_trial_results}
"""
\end{lstlisting}

\subsection{Placeholder in Prompts} \label{sec:promptPlaceholder}
The prompts contain multiple placeholders. The definition of each value is shown in Table \ref{tab:promptPlaceholder}. These values are dynamically set at runtime, substituted into the prompt, and passed to the LLMs.

\input{tables/prompt_placeholder}

\section{Hyperparameters}\label{sec:hyperparameters}
In our developed agent system, all seven types of hyperparameters are set at runtime. Table \ref{tab:hyperparams} summarizes their overview and the values used in this study.

\input{tables/hyper_parameters}

%% file: ideas/best_feature_map_formula.tex
\newcommand{\contshift}{\hspace{-1.0em}}

\begin{tcolorbox}[width=\dimexpr\linewidth-10pt\relax,colback=white, boxrule=0.5pt,
                  left=2mm, right=2mm, top=1mm, bottom=1mm, before skip=0pt, after skip=0pt]
\begin{align*}
\ket{\Phi(x)}
&= \text{MultiRZ}\Biggl(
    \pi\,\sum_{l=1}^{5}\Delta_{l}\,x_{16(l-1)+10}
   \Biggr)
   \prod_{l=1}^{5}\Biggl[
     \bigotimes_{j=0}^{9} R_Y\Bigl(\pi\,x_{16(l-1)+j}\Bigr) \\
&\contshift\qquad\cdot \prod_{j=0}^{9}
       \text{CRX}^{(l)}_{j,(j+1)\,\mathrm{mod}\,10}\Biggl(
         \pi\Bigl(
           0.5\,x_{16(l-1)+10+(j\,\bmod 6)}
           +0.5\,x_{16(l-1)+10+((j+1)\,\bmod 6)} \\
&\contshift\qquad\qquad\qquad\quad
           +0.1\bigl(
             x_{16(l-1)+10+(j\,\bmod 6)}
            -x_{16(l-1)+10+((j+1)\,\bmod 6)}
           \bigr)
         \Bigr)
       \Biggr) \\
&\contshift\qquad\cdot \prod_{j=0}^{9}
       \text{CRY}^{(l)}_{j,(j+2)\,\mathrm{mod}\,10}\Biggl(
         \pi\,\frac{
           x_{16(l-1)+10+(j\,\bmod 6)}
          +x_{16(l-1)+10+((j+2)\,\bmod 6)}
          +x_{16(l-1)+10+((j+4)\,\bmod 6)}
         }{3}
       \Biggr) \\
&\contshift\qquad\cdot \prod_{j=0}^{9}
       \text{CRot}^{(l)}_{j,(j+3)\,\mathrm{mod}\,10}\Biggl(
         \pi\,\lambda_{l}\,
         \Bigl(
           0.5\,x_{16(l-1)+10+(j\,\bmod 6)}
          +0.5\,x_{16(l-1)+10+((j+1)\,\bmod 6)}
         \Bigr)
       \Biggr) \\
&\contshift\qquad\cdot \prod_{j=0}^{9}
       \text{ISWAP}_{j,(j+5)\,\mathrm{mod}\,10}\Bigl(
         \pi\,\gamma
       \Bigr)
   \Biggr]\,
   \ket{0}^{\otimes 10}
\end{align*}

\end{tcolorbox}

%% file: circuits/best_circuit.tex
\begin{figure}[ht]
\centering

\makebox[\linewidth][c]{
\scalebox{0.8}{
\begin{quantikz}[row sep = {0.6cm,between origins}, column sep=0.2cm, thin lines]
    \lstick{$\ket{0}$} & \gate{RY}\gategroup[wires=10,steps=22,
            style={dashed,rounded corners,fill=gray!10,
                   inner xsep=1pt,inner ysep=2pt},
            background,
            label style={label position=above, yshift=0.2cm}
        ]{$\times 5$} & \ctrl{1} & \qw & \qw & \qw & \qw & \qw & \qw & \qw & \qw & \gate{RX} & \ctrl{2} & \qw & \qw & \qw & \qw & \qw & \qw & \qw & \qw & \gate{RY} & \qw & \qw \\
    \lstick{$\ket{0}$} & \gate{RY} & \gate{RX} & \ctrl{1} & \qw & \qw & \qw & \qw & \qw & \qw & \qw & \qw & \qw & \ctrl{2} & \qw & \qw & \qw & \qw & \qw & \qw & \qw & \qw & \gate{RY} & \qw \\
    \lstick{$\ket{0}$} & \gate{RY} & \qw & \gate{RX} & \ctrl{1} & \qw & \qw & \qw & \qw & \qw & \qw & \qw & \gate{RY} & \qw & \ctrl{2} & \qw & \qw & \qw & \qw & \qw & \qw & \qw & \qw & \qw \\
    \lstick{$\ket{0}$} & \gate{RY} & \qw & \qw & \gate{RX} & \ctrl{1} & \qw & \qw & \qw & \qw & \qw & \qw & \qw & \gate{RY} & \qw & \ctrl{2} & \qw & \qw & \qw & \qw & \qw & \qw & \qw & \qw \\
    \lstick{$\ket{0}$} & \gate{RY} & \qw & \qw & \qw & \gate{RX} & \ctrl{1} & \qw & \qw & \qw & \qw & \qw & \qw & \qw & \gate{RY} & \qw & \ctrl{2} & \qw & \qw & \qw & \qw & \qw & \qw & \qw \\
    \lstick{$\ket{0}$} & \gate{RY} & \qw & \qw & \qw & \qw & \gate{RX} & \ctrl{1} & \qw & \qw & \qw & \qw & \qw & \qw & \qw & \gate{RY} & \qw & \ctrl{2}& \qw & \qw & \qw & \qw & \qw & \qw \\
    \lstick{$\ket{0}$} & \gate{RY} & \qw & \qw & \qw & \qw & \qw & \gate{RX} & \ctrl{1} & \qw & \qw & \qw & \qw & \qw & \qw & \qw & \gate{RY} & \qw & \ctrl{2} & \qw & \qw & \qw & \qw & \qw \\
    \lstick{$\ket{0}$} & \gate{RY} & \qw & \qw & \qw & \qw & \qw & \qw & \gate{RX} & \ctrl{1} & \qw & \qw & \qw & \qw & \qw & \qw & \qw & \gate{RY} & \qw & \ctrl{2} & \qw & \qw & \qw & \qw \\
    \lstick{$\ket{0}$} & \gate{RY} & \qw & \qw & \qw & \qw & \qw & \qw & \qw & \gate{RX} & \ctrl{1} & \qw & \qw & \qw & \qw & \qw & \qw & \qw & \gate{RY} & \qw & \qw & \ctrl{-8} & \qw & \qw \\
    \lstick{$\ket{0}$} & \gate{RY} & \qw & \qw & \qw & \qw & \qw & \qw & \qw & \qw & \gate{RX} & \ctrl{-9} & \qw & \qw & \qw & \qw & \qw & \qw & \qw & \gate{RY} & \qw & \qw & \ctrl{-8} & \qw
\end{quantikz}
}
}

\vspace{1em}

\makebox[\linewidth][c]{
\scalebox{0.8}{
\begin{quantikz}[row sep = {0.6cm,between origins}, column sep=0.2cm, thin lines]
    \qw \gategroup[wires=10,steps=20,
            style={dashed,rounded corners,fill=gray!10,
                   inner xsep=1pt,inner ysep=2pt},
            background,
            label style={label position=above, yshift=0.2cm}]{} & \ctrl{3} & \qw & \qw & \qw & \qw & \qw & \qw & \gate{Rot} & \qw & \qw & \gate[wires=6]{XY} & \qw & \qw & \qw & \qw & \qw & \qw & \qw & \qw & \qw & \gate[wires=10]{RZ} & \qw \\
    \qw & \qw & \ctrl{3} & \qw & \qw & \qw & \qw & \qw & \qw & \gate{Rot} & \qw & \qw & \qw & \gate[wires=6]{XY} &  & \qw & \qw & \qw & \qw & \qw & \qw & \qw & \qw \\
    \qw & \qw & \qw & \ctrl{3} & \qw & \qw & \qw & \qw & \qw & \qw & \gate{Rot} & \qw & \qw & \qw & \qw & \gate[wires=6]{XY} & \qw & \qw & \qw & \qw & \qw & \qw & \qw \\
    \qw & \gate{Rot} & \qw & \qw & \ctrl{3} & \qw & \qw & \qw & \qw & \qw & \qw & \qw & \qw & \qw & \qw & \qw & \qw & \gate[wires=6]{XY} & \qw & \qw & \qw & \qw & \qw & \qw \\
    \qw & \qw & \gate{Rot} & \qw & \qw & \ctrl{3} & \qw & \qw & \qw & \qw & \qw & \qw & \qw & \qw & \qw & \qw & \qw & \qw & \qw & \gate[wires=6]{XY} & \qw & \qw & \qw & \qw \\
    \qw & \qw & \qw & \gate{Rot} & \qw & \qw & \ctrl{3} & \qw & \qw & \qw & \qw & \ghost{XY} & \qw & \qw & \qw & \qw & \qw & \qw & \qw & \qw & \qw & \qw & \qw & \qw \\
    \qw & \qw & \qw & \qw & \gate{Rot} & \qw & \qw & \ctrl{3} & \qw & \qw & \qw & \qw & \ghost{XY} & \qw & \qw & \qw & \qw & \qw & \qw & \qw & \qw & \qw & \qw & \qw \\
    \qw & \qw & \qw & \qw & \qw & \gate{Rot} & \qw & \qw & \ctrl{-7} & \qw & \qw & \qw & \qw & \ghost{XY} & \qw & \qw & \qw & \qw & \qw & \qw & \qw & \qw & \qw & \qw \\
    \qw & \qw & \qw & \qw & \qw & \qw & \gate{Rot} & \qw & \qw & \ctrl{-7} & \qw & \qw & \qw & \qw & \ghost{XY} & \qw & \qw & \qw & \qw & \qw & \qw & \qw & \qw & \qw \\
    \qw & \qw & \qw & \qw & \qw & \qw & \qw & \gate{Rot} & \qw & \qw & \ctrl{-7} & \qw & \qw & \qw & \qw & \ghost{XY} & \ghost{RZ} & \qw & \qw & \qw & \qw & \qw & \qw & \qw
\end{quantikz}
}
}

\caption{The circuit diagram of the best performing quantum feature map.}
\label{fig:best_circuit}
\end{figure}

%% file: tables/prompt_placeholder.tex
\begin{table}
\caption{Placeholders used in our agent system.}
\label{tab:promptPlaceholder}
\setlength{\tabcolsep}{4pt}
\begin{ruledtabular}
\begin{tabular}{lll}
Component & Name & Description \\
\hline
Generation & \texttt{\{max\_trial\_num\}}
  & Maximum number of experimental trials \\
 & \texttt{\{current\_trial\}}
  & Index of the current trial \\
 & \texttt{\{device\_n\_qubit\}}
  & Number of qubits available on the quantum device \\
 & \texttt{\{idea\_num\}}
  & Maximum number of ideas generated simultaneously per trial \\
 & \texttt{\{previous\_trial\}}
  & Index of the previous trial \\
 & \texttt{\{review\_comment\}}
  & Review comment from the previous trial \\
 & \texttt{\{few\_shot\_examples\}}
  & Few-shot examples for idea scoring (see Listing~\ref{lst:IdeaScoringFewShots}) \\
 & \texttt{\{current\_round\}}
  & Index of the current scoring round \\
 & \texttt{\{max\_scoring\_round\}}
  & Maximum number of scoring rounds per idea \\
 & \texttt{\{related\_work\}}
  & Summary of papers retrieved from the database \\
 & \texttt{\{max\_summary\_words\}}
  & Maximum number of words in the summary per paper \\
 & \texttt{\{raw\_content\}}
  & Full text of the paper to be summarized \\
 & \texttt{\{max\_reflection\_round\}}
  & Maximum number of reflection rounds per trial \\
 & \texttt{\{previous\_idea\}}
  & Idea generated in the previous trial \\
 & \texttt{\{previous\_score\}}
  & Score of the idea from the previous trial \\
 & \texttt{\{code\}}
  & Template for the implementation code \\
 & \texttt{\{idea\}}
  & Ideas generated in the current trial \\
 & \texttt{\{pennylane\_operations\}}
  & Available PennyLane operations and their descriptions \\
Validation & \texttt{\{methods\}}
  & PennyLane methods used in the generated code \\
 & \texttt{\{references\}}
  & Documentation of the methods retrieved from the database \\
 & \texttt{\{error\_messages\_string\}}
  & Error messages encountered during program validation \\
 & \texttt{\{warning\_messages\_string\}}
  & Warning messages encountered during program validation \\
Review & \texttt{\{max\_suggestion\_num\}}
  & Maximum number of improvement suggestions \\
 & \texttt{\{last\_trial\_num\}}
  & Trial number of the idea under review \\
 & \texttt{\{last\_trial\_results\}}
  & Generated ideas and their evaluation results \\
 & \texttt{\{quantum\_gate\_list\}}
  & List of quantum gates available in the program \\
\end{tabular}
\end{ruledtabular}
\end{table}

%% file: tables/hyper_parameters.tex
\begin{table}[ht]
\caption{Hyperparameters used in our agent system.}
\label{tab:hyperparams}
\setlength{\tabcolsep}{4pt}
\begin{ruledtabular}
\begin{tabular}{lllr}
Component & Name & Description & Value \\
\hline
All & n\_qubits
  & Number of qubits available on the quantum device
  & 10 \\
& max\_trial\_num
  & Maximum number of experimental trials
  & 30 \\
Generation & max\_idea\_num
  & Maximum number of ideas generated simultaneously per trial
  & 2 \\
& max\_scoring\_round
  & Maximum number of scoring rounds per idea
  & 3 \\
& max\_reflection\_round
  & Maximum number of reflection rounds per trial
  & 3 \\
& max\_paper\_per\_query
  & Maximum number of papers retrieved per search query
  & 3 \\
Review & max\_suggestion\_num
  & Maximum number of improvement suggestions during idea review
  & 3 \\
\end{tabular}
\end{ruledtabular}
\end{table}